\documentclass[structabstract]{aa}
\usepackage{graphicx}
\usepackage{txfonts}
\usepackage{natbib}
\bibpunct{(}{)}{;}{a}{}{,} 

\newcommand{\micron}{$\mu$m}

\def\N2H{N$_2$H$^+$}

\def\apjs{ApJS}
\def\apjl{ApJL}

\def\s{$\pm$}

\def\s12{\mbox{$S_{\rm 1.2mm}$}}
\def\micron{$\mu$m}
\def\la{\mathrel{\mathchoice {\vcenter{\offinterlineskip\halign{\hfil
$\displaystyle##$\hfil\cr<\cr\sim\cr}}}
{\vcenter{\offinterlineskip\halign{\hfil$\textstyle##$\hfil\cr
<\cr\sim\cr}}}
{\vcenter{\offinterlineskip\halign{\hfil$\scriptstyle##$\hfil\cr
<\cr\sim\cr}}}
{\vcenter{\offinterlineskip\halign{\hfil$\scriptscriptstyle##$\hfil\cr
<\cr\sim\cr}}}}}
\def\ga{\mathrel{\mathchoice {\vcenter{\offinterlineskip\halign{\hfil
$\displaystyle##$\hfil\cr>\cr\sim\cr}}}
{\vcenter{\offinterlineskip\halign{\hfil$\textstyle##$\hfil\cr
>\cr\sim\cr}}}
{\vcenter{\offinterlineskip\halign{\hfil$\scriptstyle##$\hfil\cr
>\cr\sim\cr}}}
{\vcenter{\offinterlineskip\halign{\hfil$\scriptscriptstyle##$\hfil\cr
>\cr\sim\cr}}}}}

\newcommand{\beq}{\begin{equation}}
\newcommand{\eeq}{\end{equation}}
\newcommand{\bdi}{\begin{displaymath}}
\newcommand{\edi}{\end{displaymath}}

\begin{document}


\title{ On the shape of the mass-function of dense clumps in the Hi-GAL fields }

\subtitle{I. SED determination and global properties of the mass-functions}

\author{
L. Olmi \inst{\ref{inst1},\ref{inst2}} \and 
D. Angl{\'e}s-Alc{\'a}zar\inst{3} \and 
D. Elia\inst{4} \and 
S. Molinari\inst{4} \and 
L. Montier\inst{5} \and 
M. Pestalozzi\inst{4} \and 
S. Pezzuto\inst{4} \and 
D. Polychroni\inst{4} \and 
I. Ristorcelli\inst{5} \and 
J. Rodon\inst{6} \and 
E. Schisano\inst{4} \and
M.D. Smith\inst{7} \and 
L. Testi\inst{8} \and
M. Thompson\inst{9} 
}

\institute{
	  INAF, Osservatorio Astrofisico di Arcetri, Largo E. Fermi 5,
          I-50125 Firenze, Italy, \email{olmi.luca@gmail.com} \label{inst1}  
\and
	  University of Puerto Rico, Rio Piedras Campus, Physics Dept., Box 23343, 
	  UPR station, San Juan, Puerto Rico (USA)  \label{inst2} 
\and
	  Department of Physics, University of Arizona, 1118 E. 4th Street, Tucson, 
          AZ 85721 (USA) \label{inst3}
\and
	  Istituto di Fisica dello Spazio Interplanetario - INAF, via Fosso del Cavaliere 100, 
	  I-00133 Roma, Italy \label{inst4}
\and
	  Centre d'Etude Spatiale des rayonnements, CNRS-UPS, 31028 Toulouse, France 
          \label{inst5} 
\and
	  Laboratoire d'Astrophysique de Marseille (UMR 6110 CNRS and Universit\'e de 
          Provence), 38 rue F. Joliot-Curie, 13388 Marseille Cedex 13, France \label{inst6} 
\and
	  School of Physical Sciences, Ingram Building, University of Kent, Canterbury, Kent, 
          CT2 7NH, United Kingdom  \label{inst7} 
\and
	  ESO, Karl-Schwarzschild-Str. 2, 85748 Garching bei München, Germany \label{inst8} 
\and      Centre for Astrophysics Research, University of Hertfordshire, College Lane,
          Hatfield, AL10 9AB, United Kingdom \label{inst9}
          }

\date{Received; accepted }


\abstract
{Stars form in dense, dusty clumps of molecular clouds, but little is known
about their origin and evolution.
In particular, the relationship between the mass distribution of these clumps
(also known as the ``clump mass function'', or CMF)
and the stellar initial mass function (IMF), is still poorly understood.}
{In order to discern the ``true'' shape of the CMF and to 
better understand how the CMF may evolve toward the IMF, large 
samples of  bona-fide pre- and proto-stellar clumps are required.
The sensitive observations of the Herschel Space Observatory (HSO)
are now allowing us to look at large clump populations
in various clouds with different physical conditions. 
}
{We analyse two fields
in the Galactic plane mapped by HSO during its science demonstration phase, as part 
of the more complete and unbiased  Herschel infrared GALactic Plane Survey (Hi-GAL).
These fields undergo a source-extraction and flux-estimation pipeline,
which allows us to obtain a sample with thousands of clumps.
Starless and proto-stellar clumps are separated using both color and positional criteria
to find those coincident with MIPS 24$\,\mu$m sources.
}
{
We describe the probability density functions of the power-law and lognormal 
models that are used to fit the CMFs, and we then find their best-fit 
parameters. For the lognormal model
we apply several statistical techniques to the data and compare their
results.
}
{
The CMFs of the two SDP fields show very similar shapes, but very different mass
scales. This similarity is confirmed by the values of the best-fit 
parameters of either the power-law or lognormal model. The power-law model leads 
to almost identical CMF slopes, whereas the lognormal model shows that the CMFs
have similar widths.  The similar CMF shape but different mass scale 
represents an evidence that the overall process of star formation in the two regions
is very different.
When comparing with the IMF, we find that the width of the 
IMF is narrower than the measured widths of the CMF in the two SDP fields.  This may suggest 
that an additional mass selection occurs in later
stages of gravitational collapse.
}



\maketitle

\section{Introduction}
\label{sec:intro}

Stars form in dense, dusty clumps of molecular clouds, but little is known
about their origin and evolution (sometimes the term core is also used, see
Section~\ref{sec:cc}).
In particular, the relationship between the mass distribution of these clumps
(also known as the ``clump mass function'', or CMF)
and the stellar initial mass function (IMF), is poorly understood \citep{mckee2007}.
One of the reasons for this lack of understanding,
at least from the observational point of view, has been so far the difficulty 
in selecting a statistically significant sample of truly pre- and proto-stellar 
clumps from an otherwise unremarkable collection of high column density features.

Starless (or {\it pre-stellar}, if gravitationally bound) clumps
represent a very early stage of the star formation (SF)
process, before collapse results in the formation of a central protostar, and
the physical properties of these clumps can reveal important clues about their
 nature: mass, spatial distributions and lifetime are important
diagnostics of the main physical processes leading to the formation of
the clumps from the parent molecular cloud.  In addition, a comparison of the CMF 
to the IMF may help to understand what processes are responsible for further 
fragmentation of the clumps, thus determining stellar masses. Therefore, large 
samples of {\it bona-fide} starless clumps are important for comparison of 
observations with various SF models and scenarios.

Previous studies from datasets obtained with ground facilities
(e.g., \citealp{testi1998}, \citealp{motte1998}, 
\citealp{nutter2006}, \citealp{enoch2008}, 
\citealp{alves2008} and \citealp{sadavoy2010}) 
have revealed that CMFs can roughly follow either 
power-law or lognormal shapes, which in some cases closely resemble 
the observed stellar IMF. Unfortunately, these works have also emphasized the 
difficulty in discerning the form of the CMF \citep{swift2010}.
In fact, in some cases relatively small regions within
larger clouds were examined or, even when the observations produced surveys over 
larger areas, they were carried out at a single wavelength
(e.g., 850$\,\mu$m, 1.1\,mm) and used   
different set of conditions to identify ``clumps'' in molecular clouds. 

This scenario changed recently thanks to submillimeter continuum surveys
based on telescopes placed on (sub)orbital platforms, namely the 
{\it Balloon-borne Large Aperture Submillimeter Telescope}
(BLAST; \citealp{pascale2008}) and the {\it Herschel Space Observatory} (HSO).
BLAST carried out simultaneous observations at 
$\lambda = 250$, 350, and 500$\,\mu$m of several Galactic star forming regions (SFRs;
\citealp{chapin2008}, \citealp{netterfield2009}, 
\citealp{olmi2009}, \citealp{roy2011}). 
The even more sensitive and higher-angular resolutions 
observations of HSO  are now allowing us to look at large clump populations 
in various clouds with different physical conditions, while using a 
self-consistent analysis to derive their physical parameters.
For example, the first results from the {\it Herschel} Gould Belt Survey
confirm that the shape of the pre-stellar CMF resembles the stellar IMF 
(\citealp{andre2010}, \citealp{konyves2010}).

In this first paper we analyse two fields 
in the Galactic plane mapped by HSO during its science demonstration 
phase (SDP). The two fields observed represent a sample of the more 
complete and unbiased {\it Herschel} infrared GALactic Plane Survey (Hi-GAL).
Hi-GAL is a key program  of HSO to carry out a 5-band photometric imaging survey 
at 70, 160, 250, 350, and $500\,\mu$m  of a $| b | \le 1^{\circ}$-wide strip of 
the Milky Way Galactic plane, originally planned for the longitude range 
$-60^{\circ} \le l \le 60^{\circ}$ \citep{molinari2010PASP}, and then extended in
subsequent proposals to the whole Galactic plane.

The two SDP fields have been thoroughly analyzed and have also been used to test
various methods of source extraction. In addition, they have different global 
properties that make them interesting for the purposes of this work. Here we use 
these two regions as a test bed for methods of analysis that will be later applied 
to the rest of the Hi-GAL survey. Therefore, the conclusions of this work should
be considered preliminary and specific for the SDP fields.

The outline of the paper is the following: in Section~\ref{sec:obs}, we give 
a general description of the Hi-GAL data.
In Section~\ref{sec:phot}, we describe the source extraction technique 
and the photometry specifically adopted in this work, while we describe 
how the spectral energy distributions (SEDs) were assembled in Section~\ref{sec:sed}. 
The statistical analysis of the CMFs is carried out in Sections~\ref{sec:FittingModels}
and \ref{sec:fitting}. We discuss our results in Section~\ref{sec:discussion} and draw
our conclusions in Section~\ref{sec:conclusions}.

\section{Observations }
\label{sec:obs}



The observations were carried out by HSO during the SDP that took place in November 2009. 
Five wavebands were simultaneously observed: the SPIRE instrument \citep{griffin2010} 
at $\lambda = 250$, 350, and 500$\,\mu$m, and the PACS instrument \citep{poglitsch2010}
at $\lambda = $70 and 160$\,\mu$m, were used (see Tab.~\ref{tab:instr}). The two observed fields were
centered at $l=30^{\circ}$ and $l=59^{\circ}$ and the final maps 
spanned $\simeq 2^{\circ}$ in both Galactic longitude and latitude. 

The detailed description of the observation settings and scanning strategy adopted 
as well as the map generation procedure is given in \citet{molinari2010AA, molinari2010PASP}.
Images of the SDP fields can be found in \citet{molinari2010AA} and an analysis of various
general properties of these regions can be found in \citet{battersby2011}, 
\citet{billot2010} and \citet{elia2010}. 
Here we summarize the most relevant SFRs known in both regions.

\subsection{The $l=30^{\circ}$ region}
\label{sec:l30}

The $\ell=30^{\circ}$ region that has been analyzed is approximately
4\,deg$^2$ in size and is dominated by the SRBY 162 (named after \citealp{solomon1987},
see also \citealp{mooney1995}; also called W43-main, see \citealp{nguyen2011})  and SRBY 171 
(also called W43-south) SFRs, with a total mass of several times 
$10^6\,M_\odot$. The clump of W43-main harbors a well-known giant HII region 
powered by a very luminous ($\sim 3.5 \times 10^6\, L_\odot$, \citealp{blum1999}
and references therein) cluster of Wolf-Rayet and OB stars. 
W43-south corresponds to a less extreme cloud, which also harbors a smaller HII region,
the well-known ultra-compact HII region G29.96$-$0.02 \citep{cesaroni1998}.

Recent analysis of the W43-main region, also in the context of the Hi-GAL project,
revealed a complex structure that could be resolved into a dense cluster of protostars, 
infrared dark clouds, and ridges of warm dust heated by high-mass stars, 
thus confirming its efficiency in forming massive stars \citep{bally2010}.
While the two SDP fields seem to have similar clustering properties \citep{billot2011},
\citet{battersby2011} show that the median temperatures and the column densities,
for all the pixels in the source masks considered, are higher
in the $\ell=30^{\circ}$ field than in $\ell=59^{\circ}$.  In addition, 
\citet{battersby2011} also speculate that the fact that the fraction of pixels with 
absorption at $8\,\mu$m in the $\ell=59^{\circ}$ field is so much lower than that 
in the $\ell=30^{\circ}$ field, could suggest that there is a lower fraction of cold, 
high-column density clouds in the $\ell=59^{\circ}$ field.

Finally, the richness of the $\ell=30^{\circ}$ region in young massive stars has also been
associated with it being approximately located at the interaction region
between one end of the Galactic bar and the Scutum spiral arm (\citealp{garzon1997}).

%
%
\begin{table}
\caption{PACS/SPIRE wavebands and beam FWHM. The measured beam FWHM for PACS at 70\,$\mu$m
is much larger than its nominal diffraction FWHM. 
}
\label{tab:instr}
\centering
\begin{tabular}{lcr}
\hline\hline
Instrument   & Band       & Beam FWHM   \\
             & [$\mu$m]   & [arcsec]    \\
\hline
PACS     & 70       & 9.2  \\
PACS     & 160      & 12.0 \\
SPIRE    & 250      & 17.0 \\
SPIRE    & 350      & 24.0 \\
SPIRE    & 500      & 35.0 \\
\hline
\end{tabular}
\end{table}

\subsection{The $\ell=59^{\circ}$ region}
\label{sec:l59}

Contrary to the $\ell=30^{\circ}$ region, the $\ell=59^{\circ}$ field is not located 
at the tip of the Galactic bar, but belongs instead to the Sagittarius 
spiral arm.  
%
%
This region covers approximately 5\,deg$^2$  and it is prominent in images
of thermal dust emission as well as in the radio and the optical
(see, e.g., \citealp{chapin2008} and \citealp{billot2010}).
The most active SFR is the Vulpecula OB association which hosts the star cluster
NGC~6823 and three bright \ion{H}{ii} regions, Sh2-86, 87 and 88.
The stellar HR diagram for NGC~6823 has been examined by \citet{massey1995}.  
They find an age of 5--7\,Myr for the bulk of the stars.  



The far-infrared (60 and 100\,\micron) emission in this region 
is dominated by several luminous high-mass 
SFRs.  Three of these have been studied by \citet{beltran2006}
and \citet{zha05} and are associated with the IRAS sources 
19368+2239, 19374+2352, and 19388+2357.  A further four IRAS regions 
(19403+2258, 19410+2336, 19411+2306, and 19413+2332) have been studied
extensively by \cite{beu02}, using CS multi-line, multi-isotopologue
observations and the 1.2\,mm dust continuum. 



%
%

\section{Source Photometry and identification of compact sources}
\label{sec:phot}

There is no standard terminology to identify the compact sources extracted in
bolometer maps by various existing algorithms. However, while ``core'' usually refers
to a smaller-scale object ($\la 0.1\,$pc), possibly corresponding to a later stage of
fragmentation, the term ``clump'' is generally used for a somewhat larger ($\ga 1\,$pc),
unresolved object, possibly composed of several cores \citep{williams2000}.
Our maps are likely a collection of both cores and clumps; however, given the distances of
the two fields being analyzed here (see Section~\ref{sec:sed} and \citealp{russeil2011}),
we think the term clump is more appropriate to refer to the compact objects extracted
in the SDP fields.
We also note that since we are constructing the clump (and not core)
mass functions, with the clumps likely being composed of smaller fragments,
no attempt will be made here to separate the gravitationally bound and unbound sources.

\subsection{Source and flux extraction in the SPIRE/PACS maps}
\label{sec:fluxextr}

As we mentioned in the introduction, the SDP fields have been useful test beds 
for various methods of source extraction and brightness estimation. 
However, for the purposes of this work it was also necessary to adopt 
a method that in the end would be able to determine source masses, and thus
the CMF, with a better accuracy compared to the original source extraction and 
brightness estimation pipeline described by \citet{elia2010} and 
\citet{molinari2011}.  This is achieved in two ways: first, the method outlined 
here defines in a consistent manner the region of emission of the {\it same volume} 
of gas/dust at different wavelengths, thus differing from the source grouping and
band-merging procedures described by \citet{molinari2011} and \citet{elia2010}. 
In addition, the SED fitting procedure is more accurate compared to that
described by \citet{elia2010} (see Section~\ref{sec:sed}).

The source extraction and brightness estimation techniques applied to the Hi-GAL maps
in this work are similar to the methods used during analysis of the BLAST05 
\citep{chapin2008} and BLAST06 data (\citealp{netterfield2009}, \citealp{olmi2009}).
However, important modifications have been applied to adapt the technique to the
SPIRE/PACS maps, as described below.

Candidate sources are identified by finding peaks after a Mexican Hat Wavelet
type convolution (MHW, hereafter; see, e.g., \citealp{barnard2004}) is 
applied to all five SPIRE/PACS maps.
Initial candidate lists from 70, 160 and $250\,\mu$m  are then found and fluxes at
all three bands extracted by fitting a compact Gaussian profile to the source.
Sources are not identified at 350 and $500\,\mu$m  due to the greater source-source and
source-background confusion resulting from the lower resolution, and also because these
two SPIRE wavebands are in general more distant from the peak of the source SED.
The Gaussian-fitted sources are then first selected based on their integrated flux 
(which cannot be lower than assigned values in each waveband) and also 
on their FWHM, which is allowed to vary from 80\% of the beam 
(to allow for pixelization effects on point sources) to 90\,arcsec.

Each temporary source list at 70, 160 and $250\,\mu$m is then purged 
of overlapping sources, by comparing the
positions of nearby sources at a given wavelength using their estimated FWHM: if two
sources are nearer than one-half of the sum of their respective FWHMs, they are taken
as being the same object. Purged lists at 70 and $160\,\mu$m  are then merged, and nearby
70 and $160\,\mu$m positions are treated as before. This merged $70/160\,\mu$m  catalog
is then compared with the (already purged) $250\,\mu$m source list, and the same 
procedure is repeated for identifying overlapping objects and merge the two catalogs. 
Thus, a final source catalog is generated that contains well separated objects detected 
from all three 70, 160 and $250\,\mu$m wavebands. 

In the next stage, Gaussian profiles are fitted again to all SPIRE/PACS maps, including 
the 350 and $500\,\mu$m wavebands, using the size and location parameters determined at 
the shorter wavelengths during the previous steps (the size of the Gaussian is convolved to 
account for the differing beam sizes).  The center of the new Gaussian fit is allowed to 
move at most by $\simeq 5$\,arcsec relative to the candidate source location, to allow for 
morphological differences at 350 and $500\,\mu$m, and for fitting one single Gaussian 
profile at those locations where more than one candidate source had been identified during the
previous steps. Then, before writing the final catalog, sources are again selected 
using their final FWHM.  Using this technique, 4702 and 2003 selected compact clumps were
identified in the $\ell=59^{\circ}$ and $\ell=30^{\circ}$ fields, respectively.

%
%
%
%
%
\begin{table*}
\caption{Median values of mass, temperature, distance and completeness limits
toward the  $\ell=30^{\circ}$ and $\ell=30^{\circ}$
fields.
}
\label{tab:median}
\centering
\begin{tabular}{lccccccccc}
\hline\hline
Population   & \multicolumn{4}{c}{\bf $\ell=30^{\circ}$ field} &
& \multicolumn{4}{c}{\bf $\ell=59^{\circ}$ field}  \\
\cline{2-5}
\cline{7-10}
& Temperature  & Mass        & Distance  & Completeness  & & Temperature  & Mass  & Distance & Completeness \\
& [K]          & [$M_\odot$] & [kpc]     & [$M_\odot$]   & & [K]       & [$M_\odot$] & [kpc] & [$M_\odot$] \\
\hline
All                 & 21.8   & 99.6  & 7.6  & 73.0   &     & 17.6   & 2.1  & 3.6  & 0.7 \\
Starless            & 19.7   & 112.6 & $-$  & $-$    &     & 17.4   & 2.2  & $-$  & $-$ \\
Proto-stellar       & 24.3   & 89.9  & $-$  & $-$    &     & 19.8   & 1.9  & $-$  & $-$ \\
\hline
\end{tabular}
\end{table*}

Monte Carlo simulations are then used to determine the completeness of
this process. Following the method outlined by \citet{netterfield2009}, 
fake sources are added to the 160 and $250\,\mu$m maps and are then processed
through the same source extraction pipeline. 
To generate these fake sources, we randomly select a fraction ($\sim 20-30$\%) 
of the sources in the final catalog, we convolve them with
the measured beam in each band and insert them back into the original maps.
The locations of these new sources are chosen to be at least 2\,arcmin from 
their original location, but not more than 4\,arcmin, so that the fake 
sources will reside in a similar background environment to their original location. 
These added sources are not allowed to overlap each other, but are not prevented 
from overlapping sources originally present in the map so that the simulation will 
account for errors due to confusion. Therefore, the sample of fake sources approximately 
reproduces the distributions in intensity and size of the original catalog.

The resulting set of maps, with both original and fake sources, is run through the 
source extraction pipeline and the extracted source parameters are compared to the 
simulation input.  The simulations are performed using smaller ($\simeq 0.2\,$deg$^2$) maps,
extracted from the original maps of the $\ell=59^{\circ}$ and $\ell=30^{\circ}$ fields
(containing a few hundreds sources),
in order to be able to run the source extraction pipeline multiple times, thus
achieving a statistical average for the mass completeness limits (estimated from the
$160\,\mu$m maps, at the 80\% confidence level), which are listed in 
Table~\ref{tab:median}.  These values have been estimated
for the median distances of each field, also listed in the same table, and typical values
of $T=20\,$K and $\beta = 2$ (the dust emissivity index, see Section~\ref{sec:sed}) 
have been used to convert the flux completeness limit into a mass completeness limit. 

\subsection{Separating starless and proto-stellar clumps}
\label{sec:mips24}

\subsubsection{Color criteria}
\label{sec:cc}

The catalogs of sources compiled with the technique described in \S\ref{sec:fluxextr}
do not attempt to separate starless and proto-stellar clumps. These populations
must be separated using independent methods, to ensure that they can be accurately
characterized. Various criteria can be found in the literature (see, e.g., \citealp{delu07}, 
\citealp{enoch2009}, \citealp{netterfield2009}, \citealp{olmi2009}), that use both
proximity criteria to infrared objects and temperature criteria. The different approaches 
may differ in their operational definition of the positional criteria and the 
identification of proto-stellar phases based on SED shape and/or other color
criteria.

In this work we combine color and positional criteria. This approach has the
advantage of being less computationally expensive, and least biased to particular
models or model parameters, as compared to a full 
SED approach (e.g., \citealp{robitaille2006}).
Given the  range of spatial resolutions and spectral coverage within the SPIRE/PACS 
wavebands, this combined approach should also be least biased as compared to 
positional criteria applied to data sets with single resolutions and limited
spectral coverage. 

Young protostars still embedded in dense clumps should peak in the far-infrared 
due to the absorption and reprocessing of light to longer wavelengths by envelopes.
Several recent studies have identified these sources using IRAC and MIPS colors 
(e.g., \citealp{harvey2006}, \citealp{jorgensen2006, jorgensen2007, jorgensen2008}).
Here, however, we also include the somewhat more restrictive criteria of 
\citet{sadavoy2010}.  In fact, since embedded protostars peak in the far-infrared, we also
require that proto-stellar clumps have strong detections at 24 and $70\,\mu$m and rising-red colors. 
The red colors will exclude stellar sources, which have flat colors in the infrared 
regime, but not extragalactic sources. Extragalactic contamination must be excluded 
separately, for example following the color criteria of \citet{gutermuth2008}.
IRAC and MIPS colors were assigned to the Hi-GAL sources using the
MIPSGAL catalog \citep{shenoy2012} and according to positional criteria
(see Section~\ref{sec:positional}). 
Our color criteria for identifying proto-stellar objects are thus as follows
(see also \citealp{sadavoy2010}):

\begin{itemize}
\item{(A)} The source flux at $24\,\mu$m has a signal-to-noise ratio (S/N)$\ge 3$ 
and the source $70\,\mu$m flux is higher than 0.1\,Jy.
\item{(B)} Source colors are dissimilar to those of star-forming galaxies 
(see \citealp{gutermuth2008}), i.e.,
\begin{eqnarray*}
\left [4.5\right ] - \left [5.8\right ] & < & \frac{1.05}{1.20} \, 
(\left [5.8\right ] - \left [8.0\right ] - 1), \,\,\, {\rm and}  \\
\left [4.5\right ] - \left [5.8\right ] & > & 1.05, \,\,\, {\rm and}  \\
\left [5.8\right ] - \left [8.0\right ] & > & 1.
\end{eqnarray*}
\item{(C1)} If the source is detected at $24\,\mu$m then,
\begin{eqnarray*}
\left [8.0 \right ] - \left [24 \right ] & > & 2.25, \,\,\, {\rm and}  \\ 
\left [3.6 \right ] - \left [5.8 \right ] & > & -0.28 \, (\left [8.0 \right ] - \left [24 \right ]) + 1.88
\end{eqnarray*}
\item{(C2)} If the source is {\it not} detected at $24\,\mu$m (i.e., due to 
the lower sensitivity), then the IRAC bands are used,
\begin{eqnarray*}
\left [3.6 \right ] - \left [5.8 \right ] & > & 1.25, \,\,\, {\rm and}  \\
\left [4.5 \right ] - \left [8.0 \right ] & > & 1.4.
\end{eqnarray*}
\end{itemize}
It should be noted, however, that sources with no $24\,\mu$m detection, and possibly
even those with no $70\,\mu$m counterpart, cannot be definitively classified as
starless, at least in a low-mass clump (see \citealp{chen2010}). These kind of
sources may in fact represent a stage intermediate between a gravitationally-bound 
starless clump (i.e., pre-stellar)  and a Class 0 protostar, and are difficult 
to identify. However, apart from these elusive objects, our color-criteria 
should be able to efficiently separate starless and proto-stellar clumps
on a statistical basis. 
Separating gravitationally bound and unbound clumps is beyond the scopes of this work.

\subsubsection{Positional criteria}
\label{sec:positional}

The positional criteria to associate a MIPSGAL counterpart to the Hi-GAL sources
could make the usual assumption that
a given clump is to be considered proto-stellar only if a young stellar object (YSO) 
candidate is found in the region of a clump where the intensity of emission is higher, 
which is generally associated with the peak submillimeter flux. However, 
this simple scenario may become more complicated in confused
regions, where several clumps are blended together, at which point the peak value could be 
off-center with respect to the volume occupied by the individual clumps.

%
%
\begin{figure}
\centering
\includegraphics[width=7.5cm,angle=0]{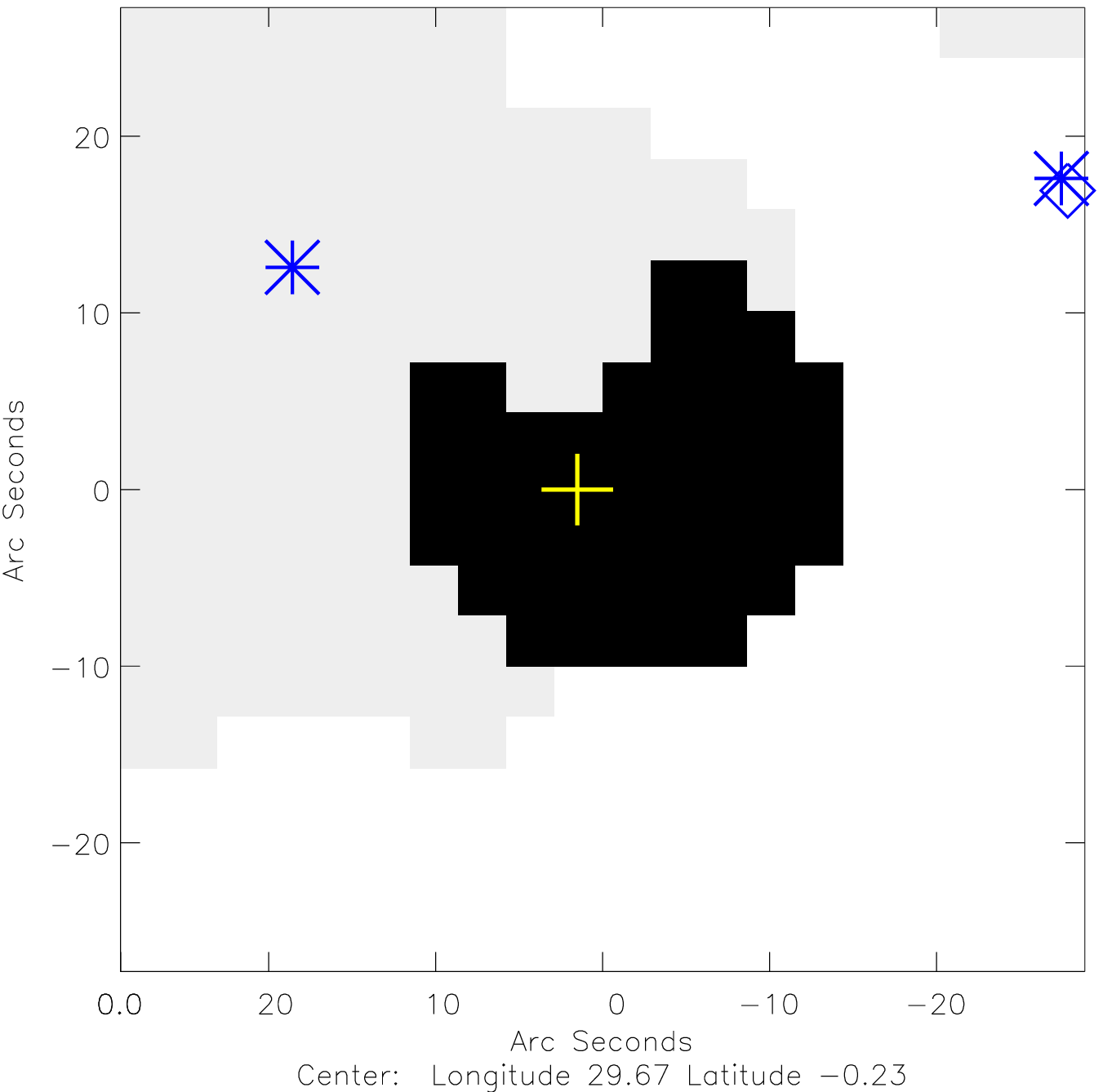}
\includegraphics[width=7.5cm,angle=0]{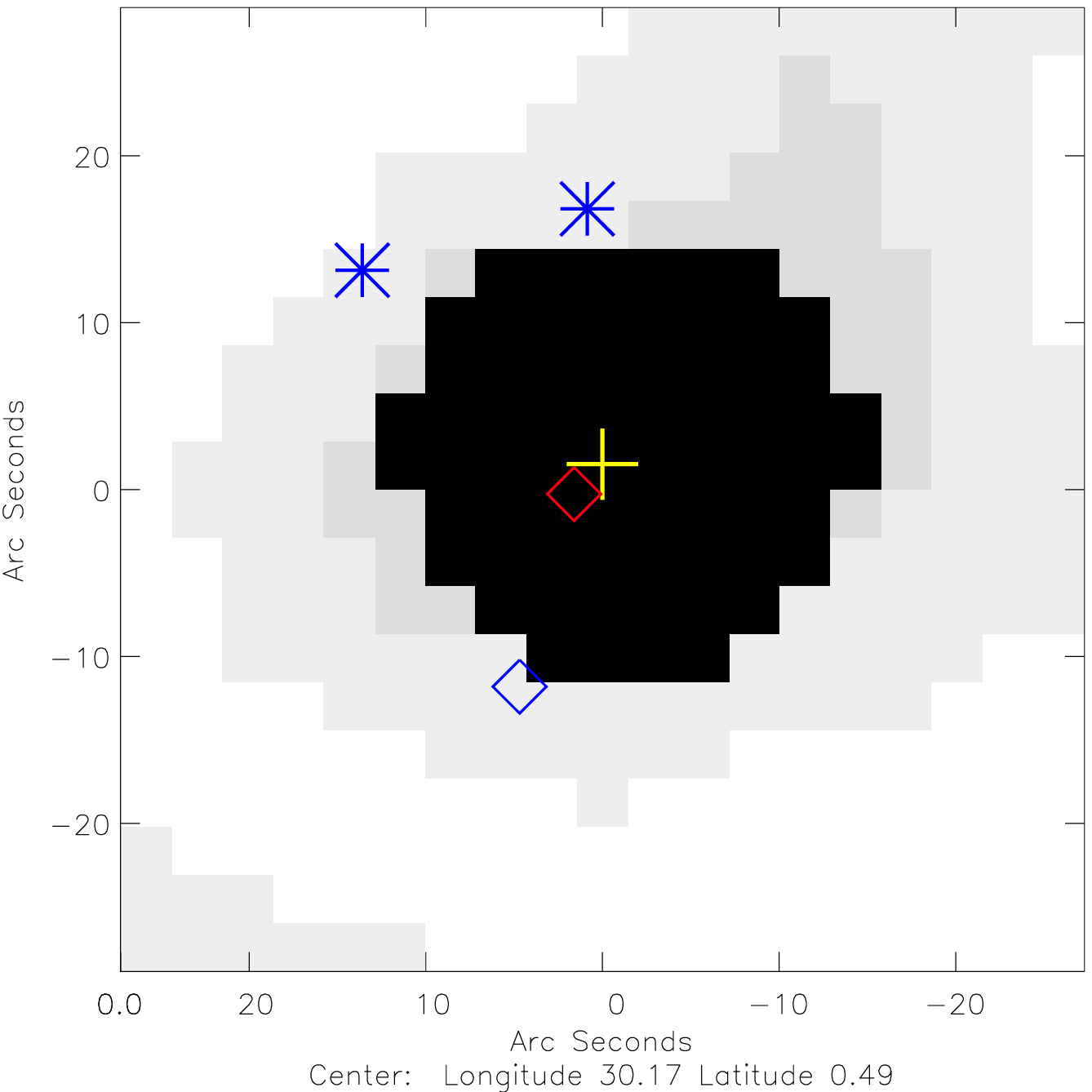}
\caption{
{\it Top.} Example of a Hi-GAL starless clump (nominal position represented with a
yellow ``+'' sign) with an irregular shape (shown by the black area, against
the gray-scale background, representing the 160\,$\mu$m emission). No MIPSGAL or MHW24
source (shown with the ``$\star$'' and ``diamond'' symbols, respectively) is falling within
the black area and thus they cannot be associated
with the Hi-GAL source.  {\it Bottom.} Example of a proto-stellar Hi-GAL
clump. One MHW24 source is associated with the submillimeter clump,
but no MIPSGAL source satisfies the coincidence and color criteria.
}
\label{fig:coincidencecriteria}
\end{figure}

The search for YSO candidates near a given submillimeter clump is further complicated
by the fact that clumps can be irregular in shape. In these cases, a circular 
approximation of the clump extent, by using the radius obtained with a 2-D Gaussian fit
to the clump intensity distribution, could lead the search algorithm to actually
probe regions beyond the ``real'' boundaries of the clump. On the other hand, a fixed
angular tolerance could clearly lead to either underestimate or overestimate the number
of YSO counterparts to the submillimeter clumps.

An alternative algorithm to ensure that the observed size and shape of the clump is considered, has 
been suggested by \citet{sadavoy2010}, in which the object location is compared to a percentage 
of the difference between the peak submillimeter intensity and the boundary intensity.
\citet{sadavoy2010} use a constant value for the boundary intensity, but in the case of the
SPIRE/PACS maps, where the clump shape and intensity has to be defined with respect to the local
background, we used a different approach.

First, a 1\,arcmin box is extracted from the PACS $160\,\mu$m flux density map, centred  
around each catalog clump. To the purpose of this procedure, we think the $160\,\mu$m map 
constitutes the best trade-off between sensitivity and angular resolution.
An average, local background level is then estimated and subtracted from the map. Intensity 
at the nominal position of the clump is evaluated in this background-subtracted map; the algorithm
also examines the nearby pixels in case the (local) peak submillimeter intensity is off-center
from the clump nominal location. The search area for coincidence with YSO candidates is then defined 
as all those pixels interior to the contour corresponding to a fraction $f=0.7$ of the peak 
intensity value found in the previous step (see Fig.~\ref{fig:coincidencecriteria}).

The fraction $f=0.7$ has been determined by trial-and-error, and thus there were always 
several cases where identifying unique YSO candidates within the specified contour 
was not always clear, particularly in crowded regions or when a dim clump was found 
near a much more intense source.
In these cases, the search area could result in an elongated or quite irregular shape, and we thus
introduced further constraints; for example, the YSO candidate must be located within a maximum
angular distance from the center of the clump, which is a function of the clump FWHM.
We note that the arbitrarity of the $f=0.7$ value may lead to some cross-contamination of the
starless and proto-stellar samples, but it should be statistically comparable in the two SDP
fields.

The presence of a Mid-Infrared source near a Hi-GAL clump was checked in two different ways.
First, the MIPSGAL catalog was used \citep{shenoy2012}. 
In order to complement this catalog, we also
decided to apply the source extraction algorithm described in \S\ref{sec:fluxextr} to the mosaicked
$24\,\mu$m MIPS maps (MIPS24, hereafter) of the $\ell=59^{\circ}$ and $\ell=30^{\circ}$ fields
(hereafter, we will refer to these MIPS24 sources as MHW24 objects). 
This approach had the advantage of being able to detect MIPS24 sources that could have been 
missed by the point response function fitting applied by the MIPSGAL team \citep{carey2009}.  
The application of the MHW method also ensured that MIPS24 sources were extracted in a 
similar and uniform fashion to the SPIRE/PACS maps.

Thus, for each Hi-GAL clump, once the contour defined by $f=0.7$ was determined, the presence
of both MIPSGAL and MHW24 sources within this contour was checked. If a
MIPSGAL source satisfies the coincidence criterion, and if in addition its colors 
satisfy the criteria described in \S\ref{sec:cc}, then this object 
is assumed to be an embedded YSO and its associated Hi-GAL clump is considered to be proto-stellar.
If a MHW24 source source satisfies the coincidence criterion, and if it also satisfies the color
criterion (A) above, then this object is equally assumed to be an embedded YSO and its 
associated Hi-GAL clump proto-stellar. Thus, if either a MIPSGAL or MHW24 source 
satisfies the color and coincidence critaria, then the associated Hi-GAL clump is 
considered proto-stellar.

\section{Submillimeter-MIR SEDs}
\label{sec:sed}

As discussed by \citet{olmi2009},
our goal here is to use a simple, single-temperature SED model to fit the
sparsely sampled photometry described in \S\ref{sec:fluxextr},
which will allow us to infer the main physical
parameters of each clump: mass, temperature and luminosity. These
quantities must be interpreted as a parameterization of a more
complex distribution of temperature and density in the clump and the
equally complex response of each instrument to these physical conditions.

%
%
\begin{figure}
\centering
\includegraphics[width=7.5cm,angle=0]{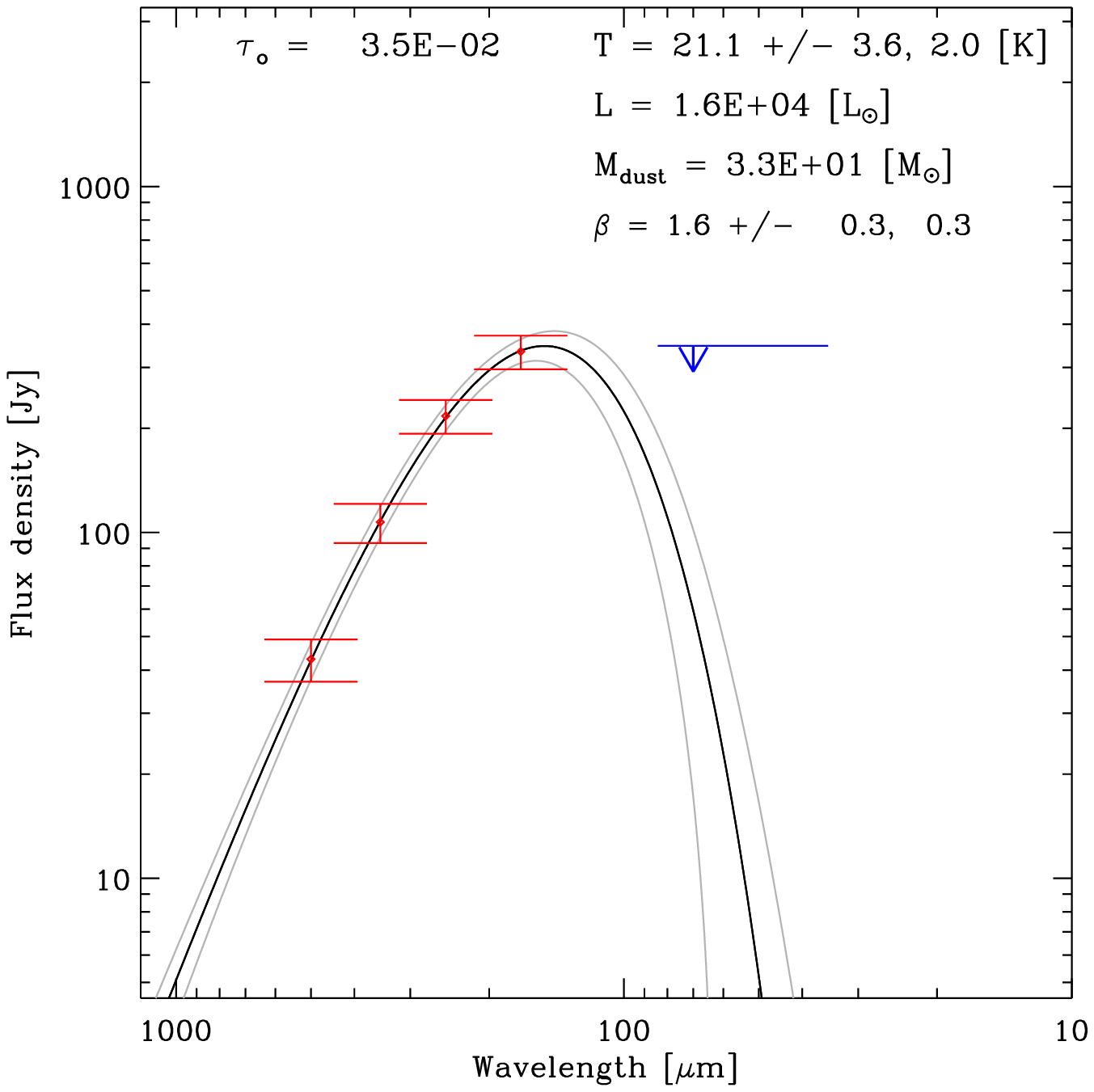}
\includegraphics[width=7.5cm,angle=0]{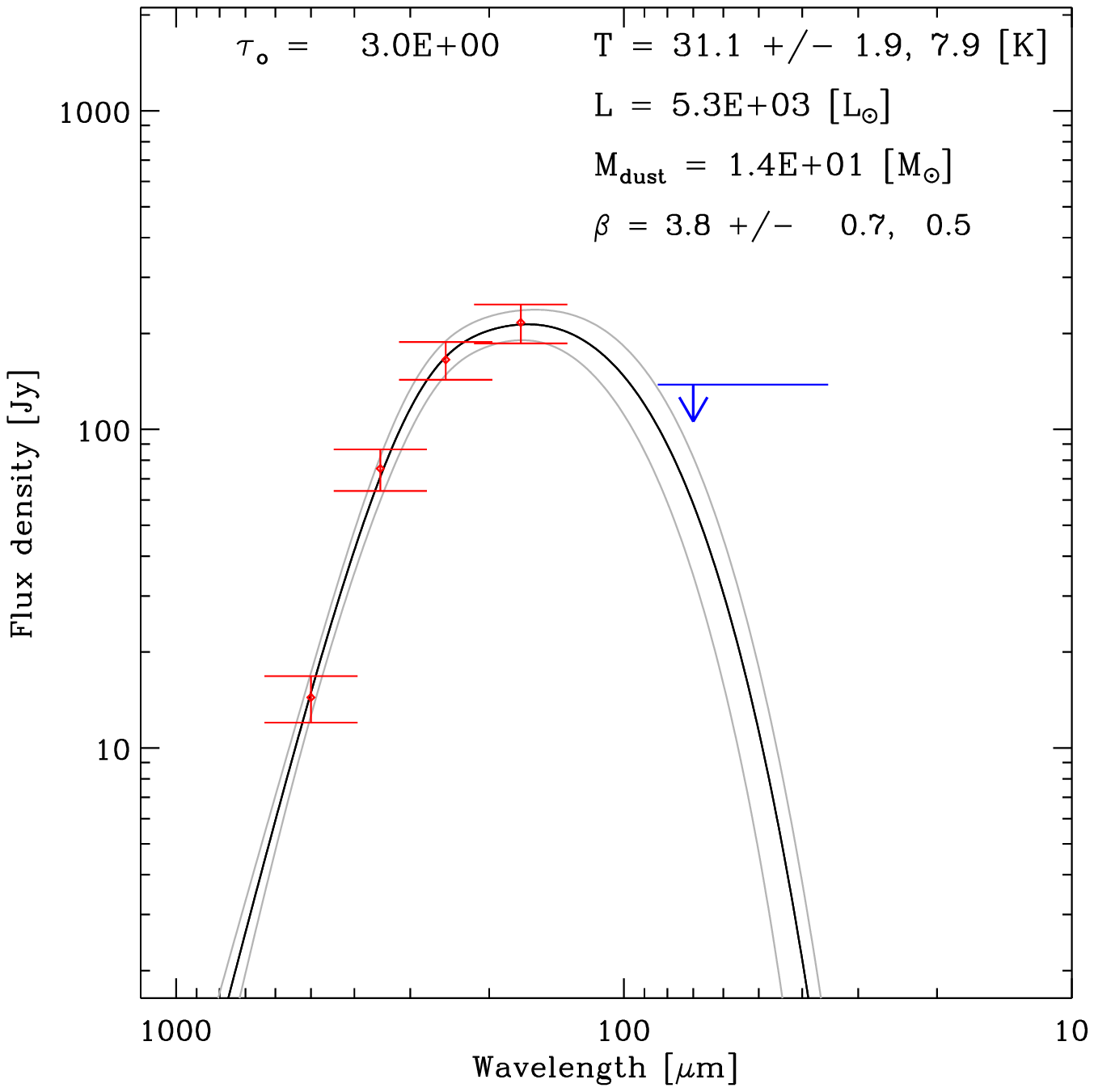}
\caption{
{\it Top.} SED of a source in the $\ell=30^{\circ}$ field.
The upper limit at $70\,\mu$m is shown and black line shows the best-fit modified blackbody,
whereas gray lines show the 68\% confidence envelope
of modified blackbody models from Monte Carlo simulations.
{\it Bottom.} Another Hi-GAL clump with a more peculiar shape
of the SED, due to the use of equation~(\ref{eq:sed}) instead of the optically-thin
approximation.
}
\label{fig:SED}
\end{figure}

In this work we mainly follow the method described by  Olmi et al. 
(2009, and references therein). Contrary to these authors, however, we do {\it not} assume
optically-thin emission and use a general isothermal modified blackbody (or gray-body)
emission of the type (see, e.g., \citealp{mezger1990}),
\begin{equation}
S_{\nu} = \Omega_{\rm s} B_{\nu}(T) \, \left [
1 - \exp ( - \tau_{\rm d} )
\right ],
\label{eq:sed}
\end{equation}
where $\tau_{\rm d}$, the dust optical depth, can be written as:
\begin{equation}
\tau_{\rm d} = \frac{A}{\Omega_{\rm s}} \left(\frac{\nu}{\nu_0}\right)^\beta
\label{eq:tau}
\end{equation}
where $A$ is a constant, $B_{\nu}(T)$ is the Planck
function, $\beta$ is the dust emissivity index, $\Omega_{\rm s}$ is the source 
angular diameter (evaluated at $160\,\mu$m) 
and the emissivity factor  is normalized at a fixed frequency $\nu_0$.
We then write the factor $A$ in terms of a total (gas + dust) clump mass,
$M$, the dust mass absorption coefficient $\kappa_0$
(evaluated at $\nu_0$), and the distance to the object, $d$:
%
\begin{equation}
A = \frac{M \kappa_0}{R_{\rm gd} d^2}.
\label{eq:mass}
\end{equation}
The distance to individual clumps was taken from \citet{russeil2011} when available 
and otherwise set to the median value (see  Table~\ref{tab:median}).
Since $\kappa_0$ refers to a dust mass, the gas-to-dust mass ratio,
$R_{\rm gd}$, is required in the denominator to infer total masses.  We adopt
$\kappa_{0} = 11$\,cm$^2$\,g$^{-1}$, evaluated at
$\nu_0=c/250\,\mu$m, and $R_{\rm gd} \simeq 100$  \citep{martin2012}.

Equation~(\ref{eq:sed}) is fit to all of the five SPIRE/PACS fluxes
(with the PACS $70\,\mu$m flux used as upper-limit, see below), 
using $\chi^2$ optimization. Color correction of the SPIRE/PACS
flux densities is performed using the filter profiles, 
and color-corrected fluxes are thus used in all subsequent applications.
Besides to $A$ and $T$, also $\beta$ and $\Omega_{\rm s}$ are allowed to vary
during the $\chi^2$ optimization, for a total of four free parameters.

As far as the parameter $\Omega_{\rm s}$ is concerned, this method
will generally result in a different value of the source size as compared
to that obtained during the source extraction procedure described in Section~\ref{sec:fluxextr}.
This is because the source extraction procedure yields a source size mainly based 
on the source shape; however, the source size delivered by the $\chi^2$ optimization 
of the SED depends solely on the source photometry. In general, we find that the latter
method tends to underestimate the source size obtained during the source extraction procedure.


Uncertainties for all model parameters are obtained from Monte Carlo simulations
(see \citealp{chapin2008}). 
Mock data sets are generated from realizations of Gaussian noise. The $\chi^2$ minimization 
process is repeated for each data set, and the resulting parameters are placed in histograms. 
Means and 68\% confidence intervals are then measured from the relevant histograms. 
Fig.~\ref{fig:SED} shows example SEDs obtained with this procedure; the
black line consists of the best-fit modified blackbody at wavelengths $> 70\,\mu$m
and the solid gray lines indicate the 68\% confidence envelope of modified blackbodies 
that fit the SPIRE and PACS data.

The $70\,\mu$m flux is always treated as an upper-limit,
because it would otherwise cause systematic deviations from
a single-temperature gray-body fit. In fact, a non-negligible fraction
of the flux at this wavelength is emitted by the warmer proto-stellar object,
either formed or in advanced stage of formation at the center of the clump, 
while most of the emission at $\lambda \ge 160\,\mu$m originates in the colder
envelope of the clump \citep{elia2010}.  We adopt ``survival analysis'' to properly 
include the upper-limit in the calculation of $\chi^2$ (see the discussion 
in \citealp{chapin2008}).

\section{ 
Description of models used to fit the CMF
}
\label{sec:FittingModels}


The goals of our analysis are to find the best fit parameters for
some selected models given the data.
In the following sub-sections we give a general description of the mathematical functions
used in this analysis, whereas in the  Appendices we outline the details of the 
numerical implementation, where all procedures were written in the 
Interactive Data Language\footnote{IDL; http://www.ittvis.com/ProductServices/IDL.aspx}.  


\subsection{Definitions}
\label{sec:defs}

For the sake of mathematical convenience we will
approximate discrete power-law and lognormal behavior with their {\it continuous}
counterparts.  Therefore, if ${\rm d}N$ represents the number of objects of mass $M$ lying between
$M$ and $M + {\rm d}M$, the number density distribution per
mass interval (or CMF), $\xi(M) = {\rm d}N/{\rm d}M$, is defined through
the relation (e.g., \citealp{chabrier2003}):
\begin{equation}
\xi(M) = \frac{{\rm d}N}{{\rm d}M} =  \frac{\xi(\log M)}{M \, \ln 10} =
\left( \frac{1}{M \, \ln 10} \right)  \frac{{\rm d}N}{{\rm d}\log M}
\label{eq:massfunct}
\end{equation}
thus, $\xi(M) {\rm d}M$ represents the number of objects with mass $M$ lying in the interval
$\left[M, M+{\rm d}M \right]$.
The  probability of a mass falling in the interval $\left[M, M+{\rm d}M \right]$
can be written for a continuous distribution as $p(M) {\rm d}M$, where $p(M)$ represents the
mass {\it probability density function} (PDF). 
The PDF and CMF must obey the following normalization conditions:
\begin{eqnarray}
& & \int_{M_{\rm inf}}^{M_{\rm sup}} p(M) {\rm d}M  =  1  \,\,\,  {\rm and} \nonumber \\
& & \int_{M_{\rm inf}}^{M_{\rm sup}} \xi(M) {\rm d}M  =   \nonumber \\
& = & \int_{\log(M_{\rm inf})}^{\log(M_{\rm sup})} \xi(\log\,M) {\rm d}\log\,M = N_{\rm tot}
\label{eq:norm}
\end{eqnarray}
where $N_{\rm tot}$ is a normalization constant which, for the case of discrete data,  can
be interpreted as the total number of objects being considered in the sample.
From Eq.~(\ref{eq:norm}) $p(M)$ can also be written as
$p(M) = \xi(M) / N_{\rm tot}$.

$M_{\rm inf}$ and $M_{\rm sup}$ denote respectively the inferior and superior
limits of the mass range for the objects in the sample,
beyond which the distribution does not follow the behavior specified by the PDF or CMF.
For example, the power-law density (see Section~\ref{sec:powerlaw}) diverges as 
$M \rightarrow 0$ so its formal distribution cannot hold for all $M \ge 0$; 
there must be some lower bound to the power-law behavior, which we denote by $M_{\rm inf}$. 
More in general, $M_{\rm inf}$ and $M_{\rm sup}$ should give us a more quantitative estimate
of the mass range where the assigned PDF gives a better description of the data.

In the following, we will also make use of the {\it complementary cumulative 
distribution function} (CCDF), which we denote $P_c(M)$ and which is defined as the
probability of the mass to fall in the interval  $\left[M, M_{\rm sup} \right]$, i.e.:
\begin{equation}
P_c(M) = \int_{M}^{M_{\rm sup}} p(M') {\rm d}M' 
\label{eq:CCDF}
\end{equation}
%

\subsubsection{Powerlaw form}
\label{sec:powerlaw}

The most widely used functional form for the CMF is the power-law: 
\begin{eqnarray}
\xi_{\rm pw}(\log\,M) & = & A_{\rm pw} \, M^{-\alpha},  \,\,\, {\rm or}  \\
\xi_{\rm pw}(M) & = &  \frac{A_{\rm pw}}{\ln 10}\,  M^{-\alpha-1}.
\label{eq:powerlaw}
\end{eqnarray}
where $A_{\rm pw}$ is the normalization constant.
The original Salpeter value for the IMF is $\alpha=1.35$ \citep{salpeter1955}.

The PDF of a power-law (continuous) distribution is given by (e.g., \citealp{clauset2009}):
\begin{equation}
p_{\rm pw}(M) = C_{\rm pw} \, M^{-\alpha-1}
\label{eq:PDFpw}
\end{equation}
where the normalization constant can be determined by applying the 
condition in Eq.~(\ref{eq:norm}), yielding:
\begin{equation}
C_{\rm pw} = \frac{ \alpha}{ (M_{\rm inf}^{-\alpha} - M_{\rm sup}^{-\alpha} ) } \,\, .
\label{eq:PDFCpw}
\end{equation}
For $\alpha > 0$ and $M_{\rm sup} \gg M_{\rm inf}$ one can use the approximation
$C_{\rm pw}  \simeq \alpha \, M_{\rm inf}^{\alpha}$ as in \citet{clauset2009} and
\citet{swift2010} (when the proper adjustments for the different definition of the
power-law exponent in Eq.~(\ref{eq:powerlaw}) are done). 
%
%

According to \citet{elmegreen1985} power-laws tend to arise when only the fragments 
can fragment, and lognormals arise when both the fragments and the interfragment gas 
can fragment during a hierarchical process of star formation. However, the power-law
functional form is also widely used because of its versatility. Past surveys of SFRs 
(see for example \citealp{swift2010} and references therein) have shown
a variety of values for $\alpha$, and the same dataset can be typically fit by one or more power-laws
with different slopes.  In particular, the power-law behavior does not extend to very low masses, 
where it displays a turnover or break below typically a few $M_\odot$.

%
%
%
%
\begin{figure*}
\centering
\includegraphics[width=9.0cm,angle=0]{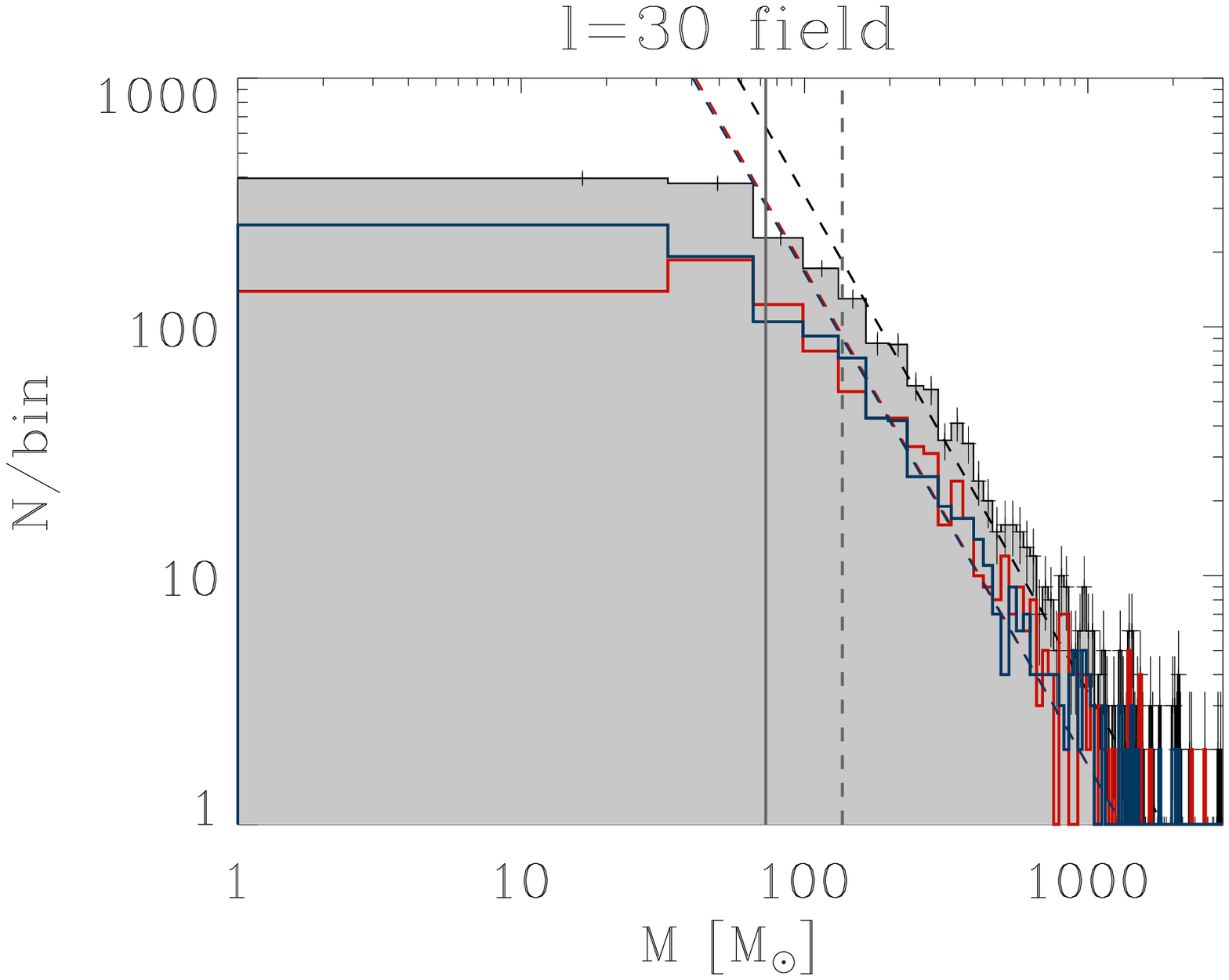}
\includegraphics[width=9.0cm,angle=0]{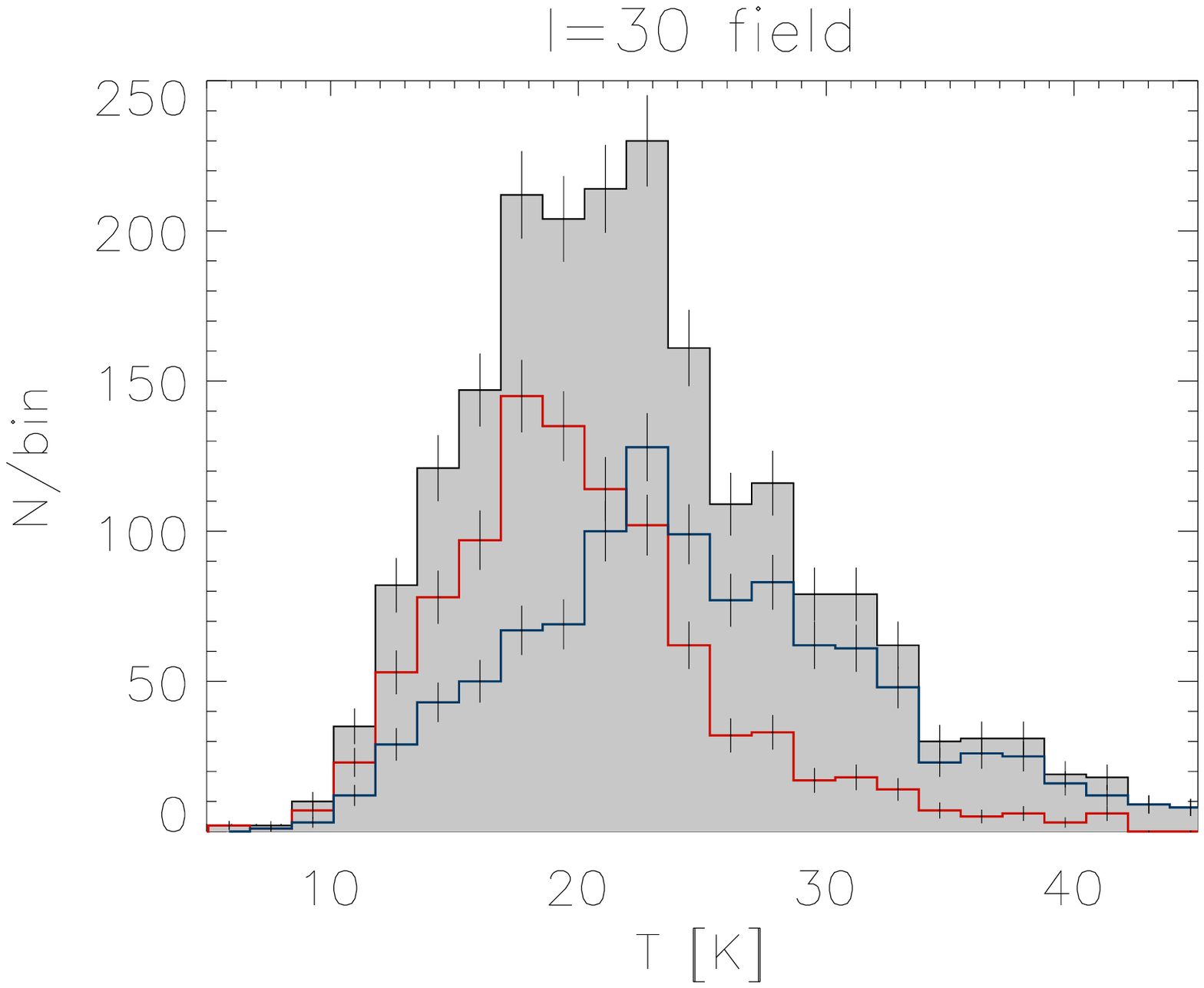} \\
\includegraphics[width=9.0cm,angle=0]{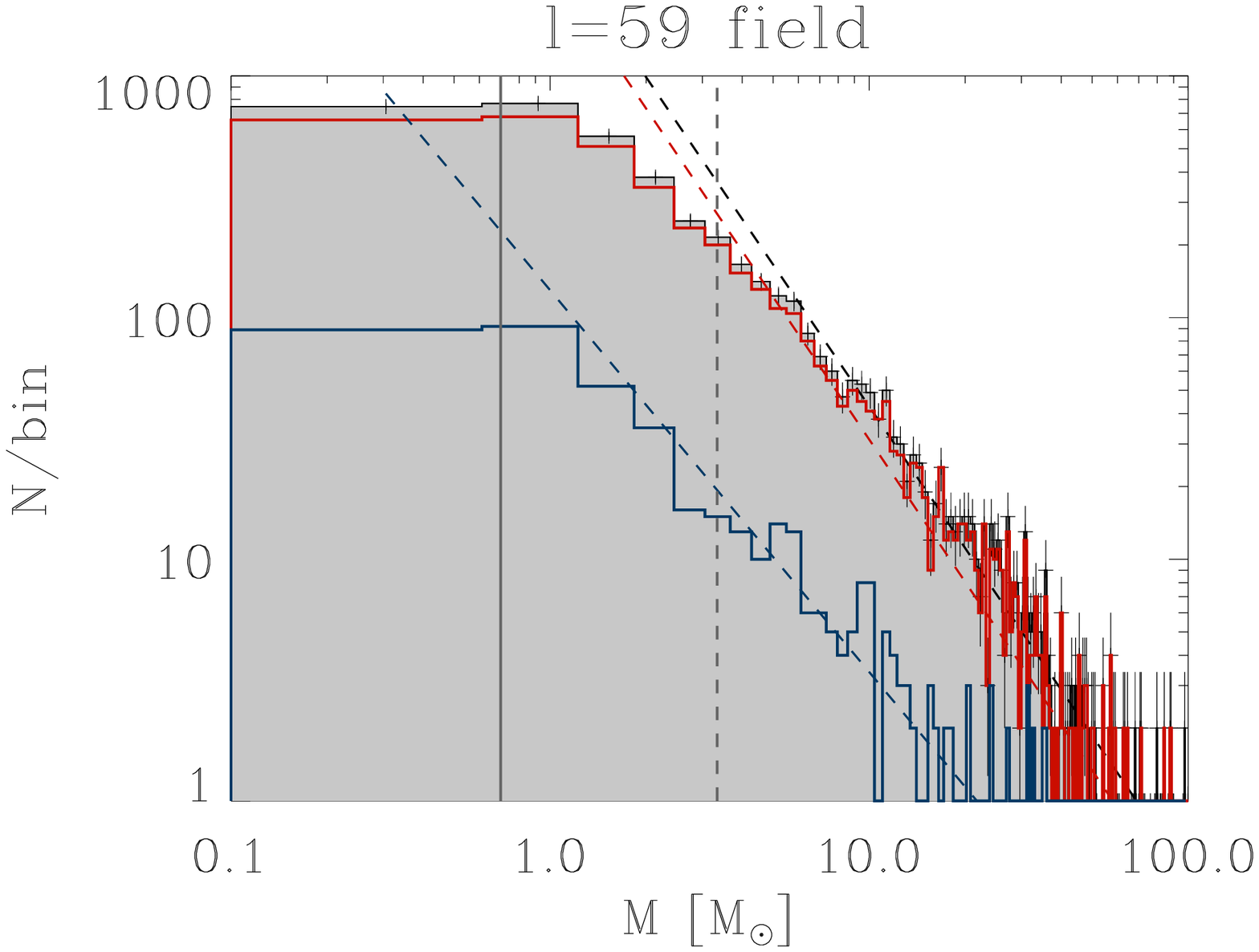}
\includegraphics[width=9.0cm,angle=0]{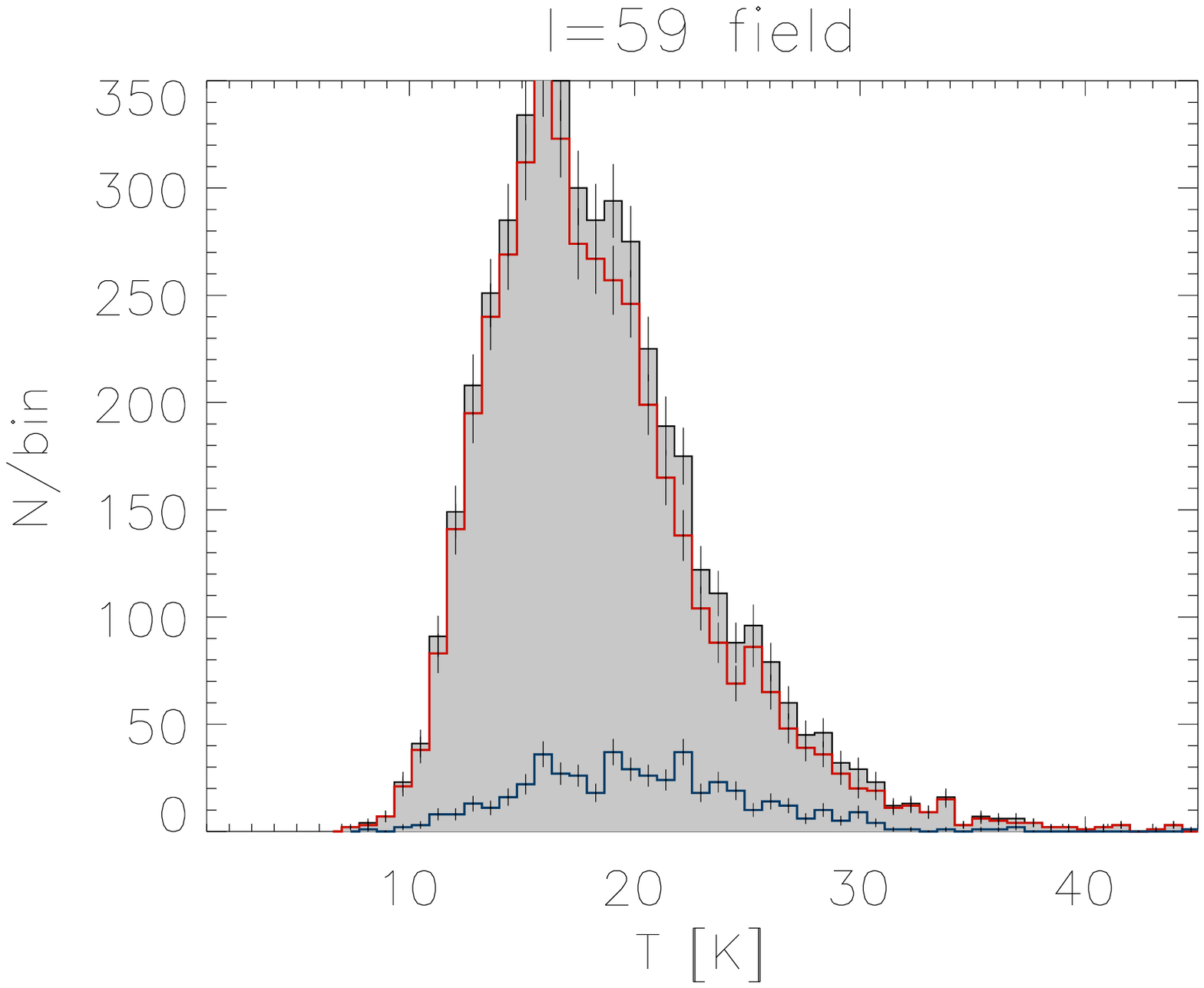} \\
\caption{
{\it Left panels.}
Log-log histograms showing the distribution of $M$ for all (black),
starless (red) and proto-stellar (blue) clumps in the $\ell=30^{\circ}$ (top)
and $\ell=59^{\circ}$ (bottom)
fields. The bin width has been chosen according to the
Freedman and Diaconis' rule (see Section~\ref{sec:lognfit}).
The dashed lines represent the results of fit to the data using the procedure PLFIT.
The vertical dashed line corresponds to the value of $M_{\rm inf}$, whereas
the solid vertical line shows the 80\% completeness limits (see Section~\ref{sec:fluxextr} and
Table~\ref{tab:median}).
Vertical bars show Poisson errors.
{\it Right panels.}
Histograms showing the distribution of clump temperature (color codes are as in the left panels)
in the $\ell=30^{\circ}$ (top) and $\ell=59^{\circ}$ (bottom)
fields. The bin width has also been chosen according to the
Freedman and Diaconis' rule.  Line styles are as before.
}
\label{fig:plfit}
\end{figure*}

%
%
%
%
%
\begin{table*}
\caption{Best-fit parameters (from PLFIT)
to the CMF of the  $\ell=30^{\circ}$ and $\ell=59^{\circ}$
fields, for the power-law distribution. The Salpeter value of the power-law exponent
for the IMF is $\alpha = 1.35$ \citep{salpeter1955}.
}
\label{tab:CMFpw}
\centering
\begin{tabular}{lccccc}
\hline\hline
Population   & \multicolumn{2}{c}{\bf $\ell=30^{\circ}$ field} &
& \multicolumn{2}{c}{\bf $\ell=59^{\circ}$ field}  \\
\cline{2-3}
\cline{5-6}
& $\alpha$  & $M_{\rm inf}$   &   & $\alpha$  & $M_{\rm inf}$ \\
&           & [$M_\odot$]     &   &           & [$M_\odot$]   \\
\hline
%
%
All                 & $1.15\pm0.15$   & $212\pm79$    &     & $1.20\pm0.15$   & $7.3\pm2.2$  \\
Starless            & $1.12\pm0.18$   & $206\pm95$    &     & $1.23\pm0.17$   & $7.5\pm2.3$  \\
Proto-stellar       & $1.06\pm0.21$   & $138\pm105$   &     & $0.58\pm0.22$   & $0.7\pm2.5$  \\
\hline
\end{tabular}
\end{table*}

\subsubsection{Lognormal form}
\label{sec:logn}

Another widely used functional form for the CMF is the {\it lognormal}, which can be 
rigorously justified because the central limit theorem applied to isothermal turbulence 
naturally produces a lognormal PDF in density. A CMF consistent with a lognormal form 
has also been observed in recent surveys of nearby SFRs (see, e.g., \citealp{enoch2006}). 
The continuous lognormal CMF can be written (e.g., \citealp{chabrier2003}):
\begin{eqnarray}
\xi_{\rm ln}(\ln\,M) = \frac{A_{\rm ln}}{\sqrt{2\pi} \, \sigma} \, 
\exp \left[ - \frac{(\ln M - \mu)^2}{2 \sigma^2} \right]
\label{eq:logn}
\end{eqnarray}
where $\mu$ and $\sigma^2 = \langle (\ln M - \langle \ln M \rangle )^2 \rangle$
denote respectively the mean mass and the variance in units of $\ln M$. 
$A_{\rm ln}$ represents a normalization constant which is evaluated in Appendix~\ref{sec:normlogn}.

The PDF of a continuous lognormal distribution can be written as (e.g., \citealp{clauset2009}):
\begin{eqnarray}
p_{\rm ln}(M) & = & \frac{C_{\rm ln}}{M} \,
\exp \left[ - \frac{(\ln M - \mu)^2}{2 \sigma^2} \right] = \\  \nonumber
& = & \frac{C_{\rm ln}}{M} \,
\exp \left[ - x^2 \right]
\label{eq:PDFln}
\end{eqnarray}
Again, by applying the normalization condition, Eq.~(\ref{eq:norm}), we find:
\begin{equation}
C_{\rm ln} = \sqrt{\frac{2}{\pi \sigma^2}  } \, \times
\left[ {\rm erfc}(x_{\rm inf}) - {\rm erfc}(x_{\rm sup}) \right]^{-1}
\label{eq:PDFCln1}
\end{equation}
where the variables $x$, $x_{\rm inf}$ and $x_{\rm sup}$ are defined in Appendix~\ref{sec:normlogn},
and we note that the parameters $M_{\rm inf}$ and $M_{\rm sup}$ are not necessarily the same
as those determined for the power-law distribution.
As we already mentioned in Section~\ref{sec:powerlaw}, if the condition $M_{\rm sup} \gg M_{\rm inf}$ 
holds, then we can write:
\begin{equation}
C_{\rm ln} \simeq \sqrt{\frac{2}{\pi \sigma^2}  } \, \times
\left[ {\rm erfc}(x_{\rm inf}) \right]^{-1}
\label{eq:PDFCln2}
\end{equation}

Then, by using the definition of PDF given in Section~\ref{sec:defs} 
and the relations for the erf and erfc functions shown in
Appendix~\ref{sec:normlogn}, the CCDF for the lognormal distribution, $P_c^{\rm ln}(M)$, can be
written as:
%
%
%
\begin{equation}
P_c^{\rm ln}(M)  = \left[ \frac{ {\rm erfc}(x) - {\rm erfc}(x_{\rm sup})  } 
{  {\rm erfc}(x_{\rm inf}) - {\rm erfc}(x_{\rm sup})  }  \right]
\label{eq:CCDFerf}
\end{equation}
%

%

\section{Fitting the data}
\label{sec:fitting}

\subsection{Evaluating the global CMF}
\label{sec:global}

In this section we apply several statistical methods to analyze the CMF of the 
two regions, and find the best-fit parameters for power-law and lognormal models. 
An important question is whether the source sample for which the CMF is constructed
should undergo more specific selection criteria, such as distance, cloud or cluster
location, etc. 
The two SDP fields may be divided into smaller sub-regions, each containing
clumps in various stages of evolution as well as already formed stars. 
Each sub-region is in turn characterized by locally different mass distributions,
which may be functions of the radial distance from the region's center (see, e.g., 
\citealp{dib2010}).  The mass function of clumps in an entire 
SDP field is thus the sum of all the sub-regions and ``local'' distributions.

The analysis of the effects of all these local distributions on the CMF of a 
larger (several sq. degrees) region is out of the scopes of this work. 
We do not attempt to separate the complete source sample of a given SDP region 
into somewhat smaller sub-samples. 
Here we limit ourselves to address the differences, if any, between the global 
CMF of the two regions and postpone the study of the aforementioned effects to 
future work.  Thus, while each source has
been assigned a specific distance, when evaluating the distance effects
between the two SDP fields we only consider the median distance of each
region (see Section~\ref{sec:distance}). 
Clearly, this also means that the CMF will be affected by
distance-dependent sensitivity and completeness effects. However,
this approach will allow us to determine if there are any significant 
differences with previous surveys, where the source sample is usually smaller 
and confined to a single cloud.

\subsection{Fitting the power-law form with the method of maximum likelihood }
\label{sec:plfit}

Various methods exist to fit a parametric model to an astronomical dataset
(see, e.g., \citealp{babu2006}). One of the most popular methods
to analyze the mass spectra of the starless and proto-stellar 
clump populations, consists of placing the masses of individual 
clumps in logarithmically spaced bins, with a lower limit on the error estimated from
the Poisson uncertainty for each bin. The resulting CMF can then be
fitted with either a power-law or a lognormal function. Then, the resulting 
best-fit slope, $\alpha$,  and the parameters of the lognormal function, 
$\mu$ and $\sigma$, usually depend somewhat on the histogram binning,
and the selected mass range, particularly when ``small'' samples of sources are used.

Alternatively, we have considered a method described by \citet{babu2006} and 
more in details by \citet{clauset2009}, that
consists of a statistically principled set of techniques that allow for the validation
and quantification of power-laws. According to \citet{clauset2009}, this method should be quite
immune to the significant systematic errors that may affect the histogram technique,
including those uncertainties associated with the histogram binning. \citet{clauset2009} have 
described a procedure that implements both the discrete and continuous maximum likelihood 
estimator (MLE) for fitting the power-law distribution to data, along with a goodness-of-fit
based approach to estimating the lower cutoff of the data. Hereafter, we will
refer to this procedure simply as ``PLFIT'', from the name of the main 
MATLAB\footnote{http://www.mathworks.it/} function performing
the aforementioned statistical operations. The results of the PLFIT method are shown in
Fig.~\ref{fig:plfit} and Table~\ref{tab:CMFpw}, and are discussed in Section~\ref{sec:discussion}.

\subsection{Fitting the lognormal form }
\label{sec:lognfit}

\begin{figure*}
\centering
\includegraphics[width=9.0cm,angle=0]{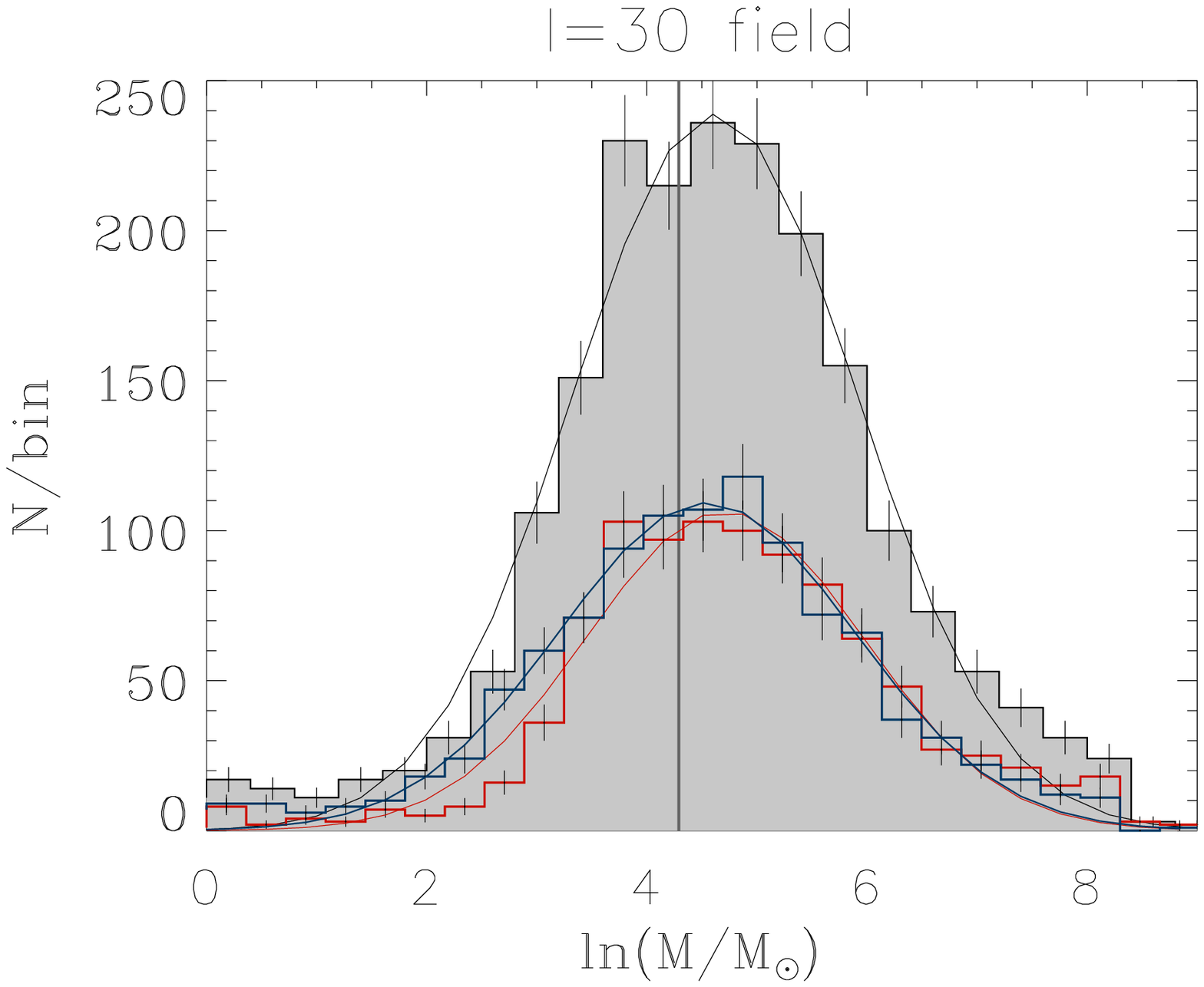}
\includegraphics[width=9.0cm,angle=0]{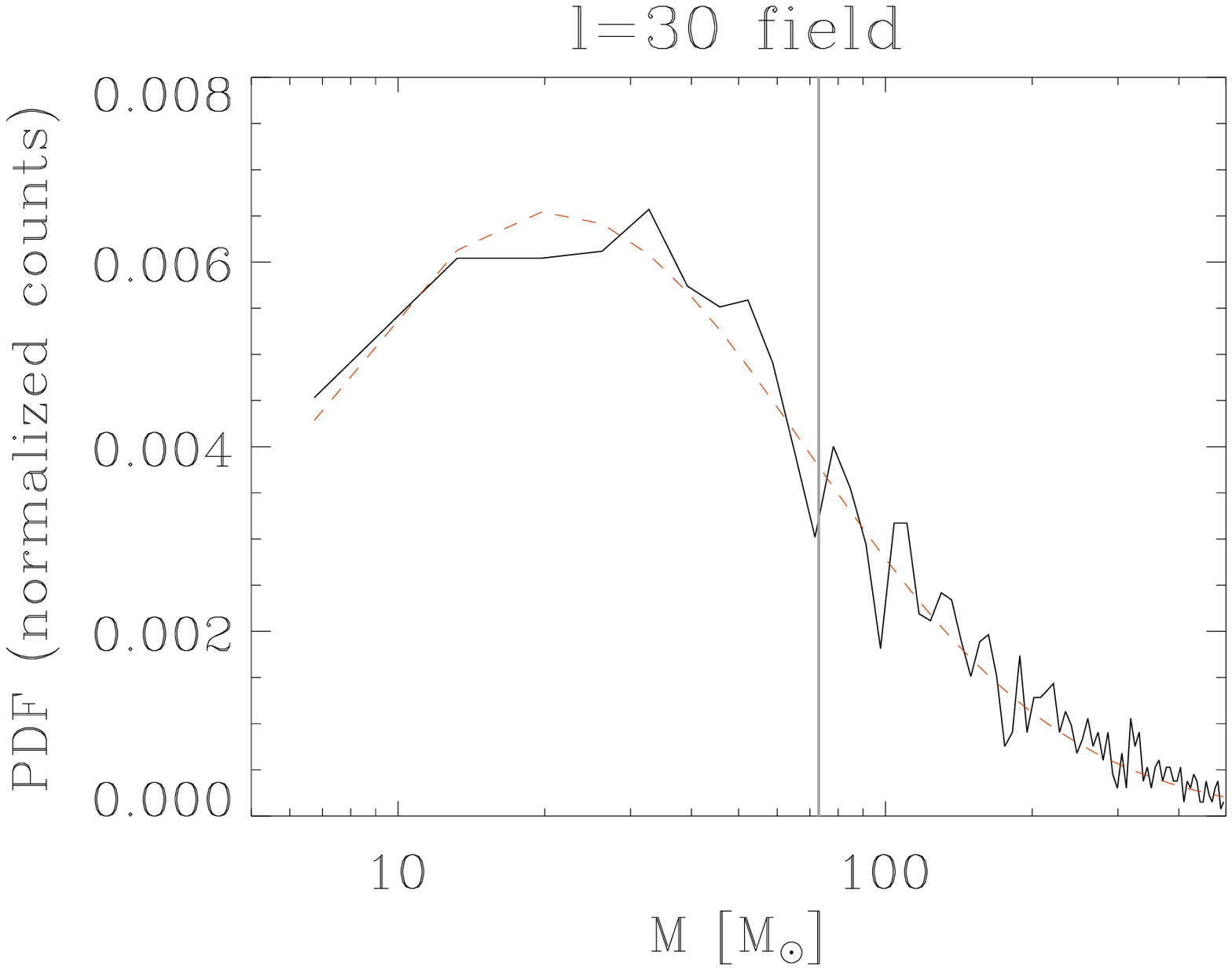}  \\
\includegraphics[width=9.0cm,angle=0]{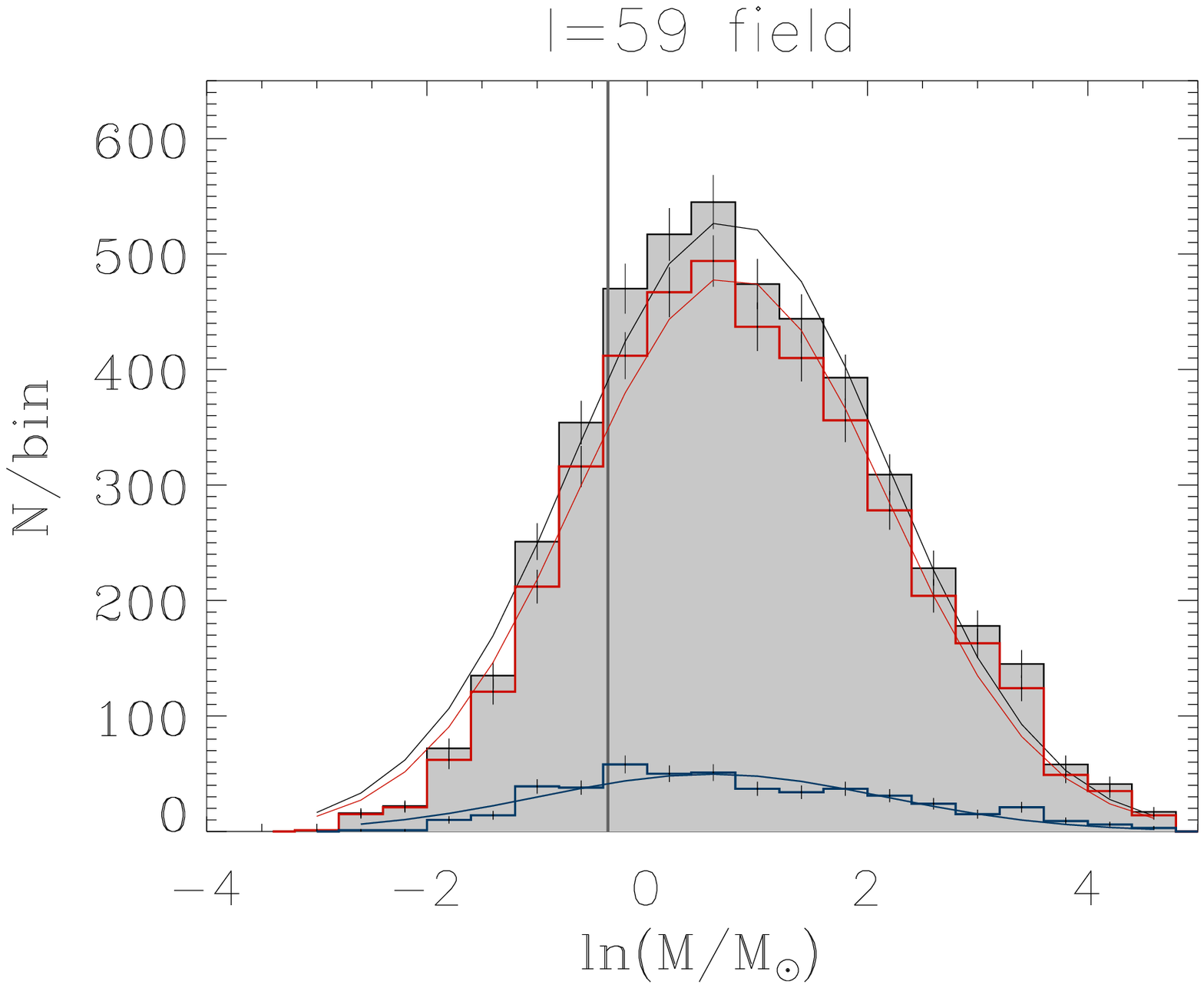}
\includegraphics[width=9.0cm,angle=0]{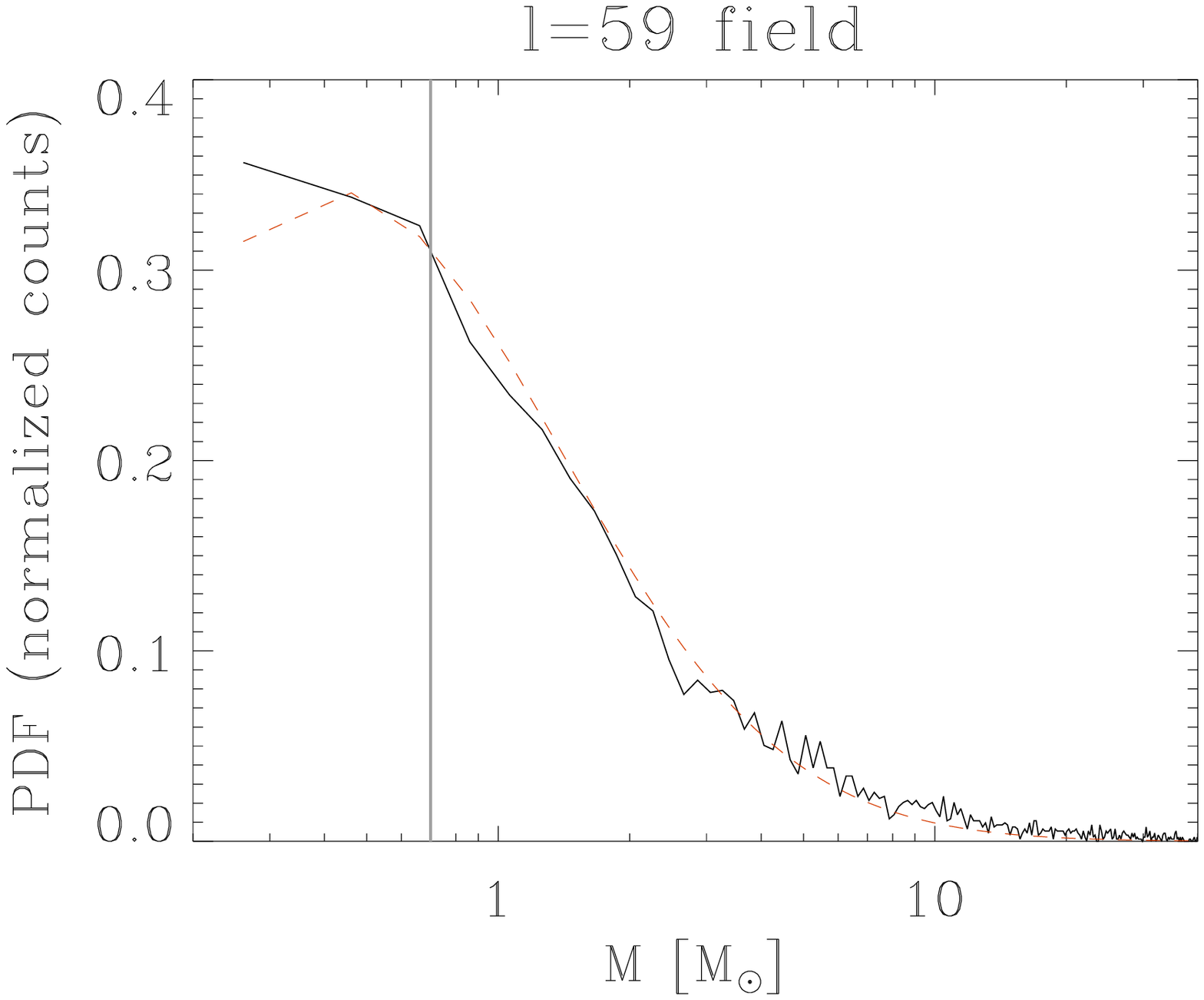}  \\
\caption{
{\it Left panels}. Histograms showing the distribution of $\ln(M)$ for all (black),
starless (red) and proto-stellar (blue) clumps in the $\ell=30^{\circ}$ (top)
and $\ell=59^{\circ}$ (bottom) fields. The solid lines represent the results of
Gaussian fits to the histograms.  The vertical solid lines represent the
completeness limits as in Fig.~\ref{fig:plfit}. Vertical bars show Poisson errors.
{\it Right panels}. PDFs of the mass distribution (black solid line) and best-fit
(red dashed line), in the $\ell=30^{\circ}$ (top)
and $\ell=59^{\circ}$ (bottom) fields.
}
\label{fig:logn}
\end{figure*}

Like Section~\ref{sec:plfit}, also in the case of the lognormal distribution we want
to avoid the uncertainties inherent in fitting data using regression models
arising from data binning.  The first method we describe here is thus based
on Bayesian regression techniques. In particular we have used 
WinBUGS\footnote{http://www.mrc-bsu.cam.ac.uk/bugs/welcome.shtml}, 
a programming language based software that is used to generate a random sample
(using Markov chain Monte Carlo, or MCMC, methods)
from the posterior distribution of the parameters of a Bayesian model.
As it is customary in Bayesian regression techniques (e.g., \citealp{gregory2005}), 
once the posterior 
distributions of the parameters of interest have been generated, they can be 
analyzed using various descriptive measures. 
In Table~\ref{tab:bayes} we show the mean values obtained for the $\mu$ 
and $\sigma$ parameters from 10000~samples MCMC runs in WinBUGS. 
We note that the procedure implemented in WinBUGS does not currently allow
to estimate the $M_{\rm inf}$ and $M_{\rm sup}$ parameters.
In the left panels of Fig.~\ref{fig:logn} we also show the histograms 
of the $\ln(M)$ values, with the solid lines representing Guassian fits 
to the histograms obtained with a standard regression technique. Both
histograms and fits are shown for graphical purposes only.

%
%
%
%

%
%
%
%
\begin{table*}
\caption{
Mean values, obtained using WinBUGS, from the posterior distributions of
the $\mu$ and $\sigma$ parameters of the lognormal distribution, shown separately for
the  $\ell=30^{\circ}$ and $\ell=59^{\circ}$ fields.
}
\label{tab:bayes}
\centering
\begin{tabular}{lccccc}
\hline\hline
Population   & \multicolumn{2}{c}{\bf $\ell=30^{\circ}$ field} &
& \multicolumn{2}{c}{\bf $\ell=59^{\circ}$ field}  \\
\cline{2-3}
\cline{5-6}
& $\mu$      & $\sigma$   &   & $\mu$      & $\sigma$  \\
& [$\ln M_\odot$] & [$\ln M_\odot$]  &  & [$\ln M_\odot$] & [$\ln M_\odot$]   \\
\hline
All                 & $4.58\pm0.03$   & $1.57\pm0.03$   &   & $0.91\pm0.02$  & $1.43\pm0.02$ \\
Starless            & $4.76\pm0.05$   & $1.50\pm0.03$   &   & $0.92\pm0.02$  & $1.42\pm0.02$ \\
Proto-stellar       & $4.41\pm0.05$   & $1.62\pm0.04$   &   & $0.89\pm0.07$  & $1.54\pm0.04$ \\
\hline
\end{tabular}
\end{table*}

While we consider the results from the Bayesian regression technique our baseline 
results, we also want to compare them 
with two alternative methods. The first one is based on the computation of
the PDF of the mass distribution, and is actually dependent on data binning. 
To compute the PDF, we plot in the right panels of Fig.~\ref{fig:logn} 
the normalized counts, i.e., the count per bin divided by the product of the 
total number of data points in the sample, $N$, and the (linear) bin width. 
For this normalization, the area  under the histogram is equal
to one, as described in Section~\ref{sec:defs}. From a probabilistic point of
view, this normalization results in a relative histogram that is most akin to the PDF.
In addition, the bin width, $W$, has been chosen according to the
Freedman and Diaconis' rule (\citealp{freedman1981})
$W = 2  (IQR)  N^{-1/3}$, where $IQR$ represents the inter-quartile range.
The best-fit results to the PDF from standard regression techniques are then 
listed in Table~\ref{tab:MLE}. 

The final method we describe here to fit the CMF with a lognormal distribution,
consists of implementing a MLE method for the lognormal function, similar to the one described
in Section~\ref{sec:plfit} for the power-law distribution. The main advantage of the
MLE technique, compared to the other methods, 
is to allow the computation of the $M_{\rm inf}$ and $M_{\rm sup}$ parameters. 
Therefore, we have numerically maximized the likelihood
of the distribution in Eq.~(\ref{eq:PDFln}) as a function of $\sigma$ and $\mu$,
using Powell's method (e.g., \citealp{press2002}). The results of this method are also
listed in Table~\ref{tab:MLE} ($4^{\rm th}$ and $5^{\rm th}$ column).
In this first instance of the MLE method for the lognormal distribution, however,
the values of $M_{\rm inf}$ and $M_{\rm sup}$ were arbitrarily fixed to
constant values.

%
%
%
%
%
\begin{table*}
\caption{Best-fit parameters to the CMF of the  $\ell=30^{\circ}$ and $\ell=59^{\circ}$ fields, for the 
lognormal distribution and the PDF and MLE methods. 
The PDF best-fits are shown in the right panels of Fig.~\ref{fig:logn}. }
\label{tab:MLE}
\centering
\begin{tabular}{lccccccccccc}
\hline\hline
Region & 
&  \multicolumn{2}{c}{\bf PDF best-fit} &
&  \multicolumn{2}{c}{{\bf MLE}\tablefootmark{a} }  & 
&  \multicolumn{4}{c}{\bf MLE with KS}  \\
\cline{3-4}
\cline{6-7}
\cline{9-12}
&
& $\mu$      & $\sigma$ &
& $\mu$      & $\sigma$ & 
& $\mu$      & $\sigma$ & $ M_{\rm inf}$  & $ M_{\rm sup}$  \\
&
& [$\ln M_\odot$] & [$\ln M_\odot$]  & 
& [$\ln M_\odot$] & [$\ln M_\odot$]  &
& [$\ln M_\odot$] & [$\ln M_\odot$]  &  [$ M_\odot$] & [$ M_\odot$]  \\
\hline
$\ell=30^{\circ}$    &  & 4.5   & 1.2  &    & 4.5   & 2.3  &  & 4.7   & 2.9  & 20   & 394 \\
$\ell=59^{\circ}$    &  & 0.54   & 1.2  &    & 1.1  & 2.6  &  & 0.53 & 1.9  & 0.5  & 10.8 \\
\hline
\end{tabular}
\tablefoot{
\tablefoottext{a}{We arbitrarily chose $M_{\rm inf} = 1\, M_\odot$ and $M_{\rm sup} = 500\, M_\odot$
for the $\ell=30^{\circ}$ region,
and $M_{\rm inf} = 0.1\, M_\odot$ and $M_{\rm sup} = 60\, M_\odot$ for $\ell=59^{\circ}$. }
}
\end{table*}

Alternatively, we may consider the values of
$M_{\rm inf}$ and $M_{\rm sup}$ to be  also unknown. These parameters can then be found by
minimizing the Kolmogorov-Smirnov (KS) statistic between the best fit model
and the data as a function of $M_{\rm inf}$ and $M_{\rm sup}$
(see Appendix~\ref{sec:lognparameters}). These latter results are listed in the 
columns $6 - 9$ of Table~\ref{tab:MLE}.

\section{Discussion}
\label{sec:discussion}


\subsection{$\ell=30^{\circ}$ field}
\label{sec:CMFpw30}

As shown in Section~\ref{sec:plfit},
because of our large sample of sources, we were able to apply the PLFIT procedure
not only to the whole sample of clumps detected toward the $\ell=30^{\circ}$ field,
but also to the starless and proto-stellar clump samples, separately.
Irrespectively of the specific sample used, we obtained $\alpha \simeq 1.1$
(see Fig.~\ref{fig:plfit} and Table~\ref{tab:CMFpw}). For $M_{\rm inf}$, 
which represents the break or turnover below which the distribution does not follow 
power-law behavior, we find $M_{\rm inf} \sim 200\, M_\odot$, except for the sample of 
proto-stellar clumps where $M_{\rm inf} = 138\, M_\odot$ but it has a large uncertainty. 

In comparing our results with previous surveys we prefer to use the results of
\citet{swift2010}, who applied similar statistical techniques 
to various (smaller) datasets from the literature. In particular, for the 
power-law functional form, we prefer not to compare our best-fit parameters with 
corresponding parameters obtained using regression models depending on data binning,
which are subject to large uncertainties, especially when the sample is relatively small 
(a few hundred sources or less).

Thus, in terms of the power-law functional form,
the estimated value of $\alpha$ agrees very well with the typical values found by
\citet{swift2010}\footnote{Their $\alpha$ values correspond to our $\alpha+1$ values.},
for both low- and high-mass SFRs. 
On the other hand, the estimated value of $M_{\rm inf} \sim 200\, M_\odot$, 
is higher compared to the values estimated by \citet{swift2010} for intermediate- 
and high-mass SFRs. 
We also note the similar values of $\alpha $ and (to a lesser extent) $M_{\rm inf}$ 
for the starless and proto-stellar clump samples.

As far as the lognormal fits are concerned, Table~\ref{tab:bayes} shows that 
the starless population in $\ell=30^{\circ}$ has a slightly higher (lower) value 
of $\mu$ ($\sigma$) compared to the proto-stellar population. More Hi-GAL fields 
need to be observed to determine whether this is a general property.
In terms of the other statistical methods, in Table~\ref{tab:MLE}
we note that while the results of the MLE and PDF methods yield $\mu$ values
quite similar to the Bayesian results, the values of $\sigma$ can differ
significantly (by almost a factor of 2).
We also note the relatively small value of $M_{\rm sup}$, 
compared to the whole range of masses in the $\ell=30^{\circ}$ field. 

It is worth noting that the completeness limit of the $\ell=30^{\circ}$ field
(see Table~\ref{tab:median}) is lower than the $M_{\rm inf}$ value.
Therefore, the peak of the $\ln(M)$ distribution in Fig.~\ref{fig:logn} is 
real, though barely constrained. Likewise, the turnover or break 
in the CMF of Fig.~\ref{fig:plfit} is also effectively observed. 
However, the fact that the turnover is not better
constrained may render all later comparisons between the power-law and lognormal 
distributions problematic.

\subsection{$\ell=59^{\circ}$ field}
\label{sec:CMFpw59}

As shown in Table~\ref{tab:CMFpw} (see also Fig.~\ref{fig:plfit})
we find that the values of $\alpha$ in the $\ell=59^{\circ}$ field
are also approximately equal for the whole sample and the starless clumps.
These values are consistent, within the errors, with those found
in the $\ell=30^{\circ}$ field, as discussed in Section~\ref{sec:distance}.
However, the $\alpha$ value for the proto-stellar clumps is indeed 
lower than that for the starless sample. Given the much smaller number
of proto-stellar clumps in the $\ell=59^{\circ}$ field, we need to
further investigate whether this could be a consequence of statistical uncertainty,
or it is instead a true property of this region.

Interestingly, the value of $M_{\rm inf}$ estimated in the $\ell=59^{\circ}$ region
is much lower than that of the $\ell=30^{\circ}$ field, as expected given that the
$\ell=59^{\circ}$ field is mostly a low- to intermediate-mass SFR. However,
contrary to the $\ell=30^{\circ}$ region, our estimated value of $M_{\rm inf}$ is 
comparable to the values estimated by \citet{swift2010} for low-mass SFRs. 
On the other hand, while \citet{swift2010} report a wide range of CMF slopes
(corresponding to $\alpha \sim 0.7 - 3.1$), our estimated values of $\alpha$ do not
appreciably vary from the $\ell=30^{\circ}$ to the $\ell=59^{\circ}$ field.

In terms of the lognormal functional form of the CMF toward 
the $\ell=59^{\circ}$ field, the results shown  in Fig.~\ref{fig:logn} and 
Tables~\ref{tab:bayes} and \ref{tab:MLE} confirm that the mass range in the
two SDP fields are quite different (see Table~\ref{tab:median}).
As in the $\ell=30^{\circ}$ region, Table~\ref{tab:MLE} shows that
the values of $\sigma$ obtained through the MLE and PDF methods can 
significantly differ  from the Bayesian results.  
We have yet to determine if this is a feature associated with the different methods.

Finally, the completeness limit 
of the $\ell=59^{\circ}$ region is  quite lower than its corresponding $M_{\rm inf}$ value.
Therefore, the peak of the $\ln(M)$ distribution in Fig.~\ref{fig:logn} is 
at least partially resolved, and the detection of a turnover in the CMF
is more reliable than it is in the $\ell=30^{\circ}$ field.

%
%
%
\begin{figure}
\centering
\includegraphics[width=9.5cm,angle=0]{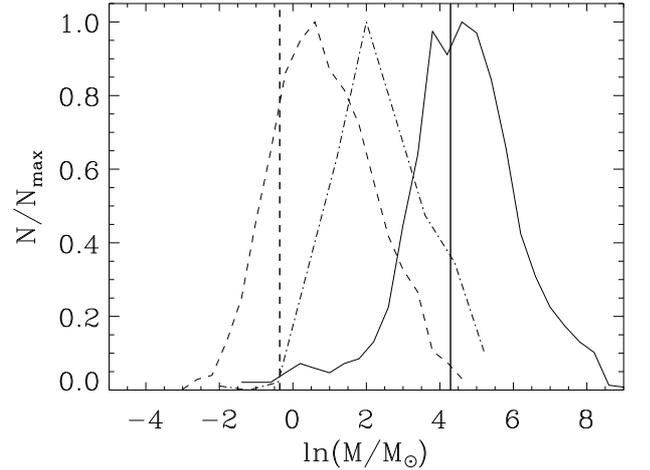}
\includegraphics[width=9.5cm,angle=0]{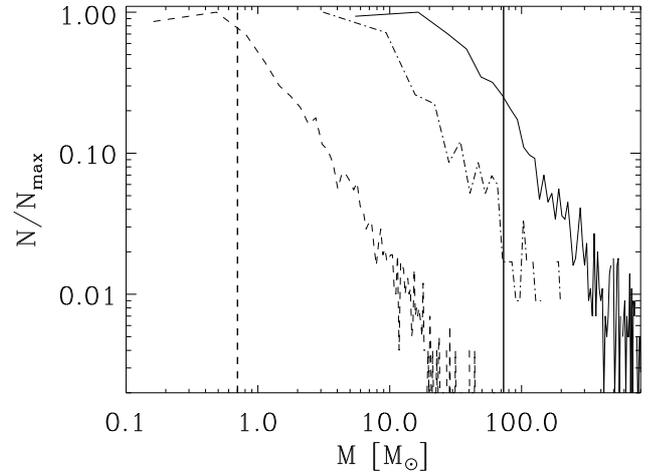}
\caption{
{\it Top.} Histogram profiles from Fig.~\ref{fig:logn}, 
with solid and dashed lines representing the $\ell=30^{\circ}$ and $\ell=59^{\circ}$ fields,
respectively. The dash-dotted line represents the distribution obtained for the
$\ell=59^{\circ}$ field traslated to the distance of $\ell=30^{\circ}$ (see text).
{\it Bottom.} CMF profiles from Fig.~\ref{fig:plfit}; line styles are as before.
The vertical lines mark the completeness limits listed in Table~\ref{tab:median} for
the $\ell=30^{\circ}$ (solid line) and $\ell=59^{\circ}$ (dashed line)  fields.
}
\label{fig:compare}
\end{figure}

\subsection{Comparing the two SDP fields }
\label{sec:compareSDP}

\subsubsection{Distance effects}
\label{sec:distance}

Comparing now the results for the $\ell=30^{\circ}$ and $\ell=59^{\circ}$ 
fields, we have already noted how the values
of $\alpha$ are similar for the two regions.
Therefore, if the evolution toward the IMF of high-mass clumps 
($\ell=30^{\circ}$) is different from those of low- and 
intermediate-mass clumps ($\ell=59^{\circ}$), it leaves
no trace on the shape of the CMF of these two regions.
This conclusion, however, is critically dependent on the correcteness
of the distance estimates, and in any case it is the result
of a survey over a large collection of different SFRs.
In fact, our findings in the $\ell=30^{\circ}$ field
are different from those of \citet{netterfield2009} who found
a different slope of the CMF for the cold and warm populations
of clumps in the Vela-C region, which is a less heterogeneous region
compared to our field.

We have also seen that the values of $\mu$ and $M_{\rm inf}$ are clearly different for 
the two regions (see Tables~\ref{tab:bayes} and \ref{tab:MLE}). 
However, and despite the uncertainty on the {\it absolute} value of $\sigma$, when comparing
the results obtained with the {\it same} regression technique in the two SDP fields,
we find that the values of $\sigma$ are remarkably similar. We can also
estimate the average values from all four methods used, and we find 
$\langle \sigma \rangle = 2.0\pm0.8$ and $1.8\pm0.6$ $[\ln\,M_\odot]$ 
for $\ell=30^{\circ}$ and $\ell=59^{\circ}$, respectively. These average values 
are also consistent with the range of $\sigma$ found by \citet{swift2010}, 
$\sim 0.7\, {\rm to}\, 3.4$, who also found the variation in the values of 
$\mu$ ($\sim -5.4\, {\rm to}\, 2.7$) larger compared to that of $\sigma$.
Therefore, the histograms representing the distribution of the $\ln(M)$ values
in the two SDP fields (Fig.~\ref{fig:logn}) are characterized by decidedly 
different mass scales, but are quite similar in shape, as it can be noted in the 
left panels of Fig.~\ref{fig:logn}.

This similarity can be noted even more clearly in the top panel of Fig.~\ref{fig:compare},
where the histogram profiles from Fig.~\ref{fig:logn}, for the $\ell=30^{\circ}$ and 
$\ell=59^{\circ}$ fields, are shown side by side.  We also compare the CMFs 
in the bottom panel of Fig.~\ref{fig:compare}.
%
%
Apart from differences due to binning, one can clearly see that the 
mass distributions are very similar, although this similarity 
should be taken as far as the completeness limits actually allow to. 

Thus, although a full comparison of the mass spectra, including the mass range below
the peak of the $\ln(M)$ distribution, will require higher sensitivities than those achieved 
in the two SDP fields, the existence of a mass scale difference between the two regions 
is clear.  It seems unlikely that the different mass scales could be due
entirely to differences in the median distance (see Tab.~\ref{tab:median}), since this
could account for a factor $\sim 4$ at most in mass sensitivity.

To test this issue more quantitatively, we have performed a simulation where the 
$\ell=59^{\circ}$ field is traslated to the same distance as the $\ell=30^{\circ}$
region. In this procedure, the angular resolution in all of the original maps 
is first degraded by convolving with a beam in each band enlarged for the 
increased distance, and then the maps are re-binned accordingly (A. Facchini, {\it priv. comm.}).
The resulting maps are run through the source extraction and SED-fitting pipelines,
and the extracted SED parameters are estimated by scaling the distance of 
each source by the ratio of the median distances in the $\ell=30^{\circ}$ and
$\ell=59^{\circ}$ fields.  

Our test thus shows that the mass 
of the sources in the traslated maps can increase due to the larger distance and, 
to a lesser extent, due to the ``merging'' of sources that are detected as separate 
objects in the original maps of the $\ell=59^{\circ}$ region.
The new mass distribution is shown in Fig.~\ref{fig:compare}, where we can clearly see
that the CMF for the traslated $\ell=59^{\circ}$ region lies between the original 
CMFs of the $\ell=59^{\circ}$ and $\ell=30^{\circ}$ fields. This appears to confirm
our earlier assumption that distance effects alone cannot explain the overall
difference in mass scales between these two specific SFRs.
We thus think that this mass scale effect is
{\it evidence that the overall process of star formation in the two regions 
must be radically different}.



\subsubsection{Clump formation efficiency}
\label{sec:cfe}

It is too early to speculate about the physical origins of the differences described in
the previous section.  In fact, we have exploited only a few percent of the Hi-GAL survey in the present analysis.
With more and more regions being analyzed we expect to find statistically significant
indications that there may indeed exist different Galactic star-forming regimes,
as stated in the Hi-GAL scientific goals.
For the present analysis, the prominence of the $\ell=30^{\circ}$ region in
star-forming indicators, compared to $\ell=59^{\circ}$ \citep{battersby2011},
including the (triggered) W43 ``mini-starburst'' complex \citep{bally2010},
and the mass scale difference we found between the two regions, could all
be related with the $\ell=30^{\circ}$ field being located near
the interaction region between one end of the Galactic bar and the
Scutum spiral arm, where high concentrations
of shocked gas are more likely to be found (\citealp{garzon1997}, \citealp{lopez1999}).

In order to further test this scenario, we have estimated 
an additional figure of merit, the clump formation 
efficiency ($CFE$), that we define as:
\beq
CFE = \frac{M_{\rm clumps} }  {M_{\rm clumps} + M_{\rm clouds} }
\eeq
where $M_{\rm clumps}$ represents the total mass of the clumps (above completeness limit)
and $M_{\rm clouds}$ is the mass of the ambient gas.
Given that the two SDP fields represent a collection of regions, possibly at
different distances, rather than being a single molecular cloud at a single distance,
two obvious problems must be considered. First off, we have to select a method 
to map the total column density, without being limited by threshold, opacity and
temperature variations in the ambient gas/dust. Second, the distance to separate parcels
of gas must be estimated.

\citet{goodman2009} have discussed and compared several methods for measuring 
column density in molecular clouds and they conclude that dust extinction is
likely the best probe. Therefore, we have downloaded the $A_{\rm v}$ extinction 
maps\footnote{http://astro.kent.ac.uk/extinction/}
obtained by  \citet{rowles2009} and \citet{froebrich2010} and used them
to estimate the total column density in the two SDP regions. Specifically,
we have selected only those pixels with $A_{\rm v} > 1$ and we have assigned 
them a distance based on the closest (within a 3\,arcmin radius) mass clump 
previously identified. Then, the extinction in each pixel has been converted
to column density using the conversion factor 
$N_{\rm H} = \beta_{\rm v} \, A_{\rm v} \,$cm$^{-2}$ 
where $\beta_{\rm v} = 2 \times 10^{21} \,$cm$^{-2}\,$mag$^{-1}$      \citep{savage1979}.
Therefore, the total mass in the clouds has been calculated as:
\beq
M_{\rm clouds} = \Delta \Omega \, \mu \, \beta_{\rm v} \, m_{\rm H2} \sum_i d_{\rm i}^2 A_{\rm v}^{\rm i} 
\eeq
where $\Delta \Omega$ is the solid angle subtended by each pixel in the extinction maps, 
$\mu = 1.38$ takes into account the cosmic He abundance, $m_{\rm H2}$ is the mass of molecular
hydrogen, and $d_{\rm i}$ and $A_{\rm v}^{\rm i}$ represent the distance and visual extinction
toward the $i-$th pixel.


Our estimated $CFE$ thus amounts to $\simeq 2.4\,$\% and $\simeq 0.7\,$\% for the 
$l=30^{\circ}$ and $l=59^{\circ}$ regions, respectively. Although the determination
of the $CFE$ is subject to several uncertainties (mainly due to the distance estimates),
this is an indication that clumps may indeed form more efficiently in the $l=30^{\circ}$
rather than in the $l=59^{\circ}$ field. However, since the relationship of the 
$CFE$ with the star formation efficiency ($SFE$) is not known, our present analysis is not
yet conclusive that the $SFE$ is also higher in the $l=30^{\circ}$ field.

\subsection{Comparing the CMF to the IMF}
\label{sec:IMF}

The qualitative similarity, observed in past studies, between the CMF and the IMF
offers support for the accepted idea that stars form from dense clumps, and thus comparing the two
distributions should allow us to learn how observed samples of clumps evolve into stars.  
This comparison is actually a complex task because CMFs are often different but the IMF
appears to be quite universal. In addition, it is difficult to understand 
whether the CMFs of different regions are intrinsically different,
or to what degree systematic differences in each dataset
(either observational or from post-processing) may contribute to the
variations seen from dataset to dataset.

\citet{swift2008} have shown that different evolutionary pathways from clumps to 
stars produce variations in the form of the resultant IMF. They also showed that
while the  power-law slope is quite robust, the width of the lognormal distribution 
is a more sensitive indicator of clump evolution. As we showed earlier, 
the average value of $\sigma$ in the two SDP fields is $\sim 1.9$ $[\ln\,M_\odot]$, 
consistent with the range of $\sigma$ found by \citet{swift2010}, and with 
an uncertainty of as much as 40\%. However, the width of the IMF has been 
measured to be narrower, between 0.3 and 0.7 $[\ln\,M_\odot]$ (e.g., \citealp{chabrier2003}).
As already suggested by \citet{swift2010} this would appear to indicate that 
an additional mass selection occurs in later stages of gravitational collapse.

\section{Conclusions}
\label{sec:conclusions}

We have analysed two fields mapped by the SPIRE and PACS instruments of HSO
during its science demonstration phase. The two fields, which are
part of the {\it Herschel} infrared GALactic Plane Survey, were
centered at $l=30^{\circ}$ and $l=59^{\circ}$ and the final maps
covered almost 10\,deg$^2$ of galactic plane.

The two regions underwent a source-extraction and flux-estimation pipeline,
which allowed us to obtain a sample with thousands of clumps.
We then applied several statistical methods to analyze the resulting CMFs, and 
found the best-fit parameters for power-law and lognormal models. No attempt was made 
to select more uniform sub-samples, except for starless and proto-stellar clumps.
Our main conclusions are the following:
\begin{itemize}
\item Our best-fit parameters for the power-law distribution show a robust slope
($\alpha \simeq 1.2$, with a $\sim\,15$\% uncertainty) when comparing the two 
SDP fields.  In contrast, we find a very different value of the 
parameter $M_{inf}$ for the two regions analyzed.
We find that $M_{\rm inf}$ is higher 
than the completeness limit in each region and is thus well defined. \\
\item We have used several statistical techniques to estimate the best-fit 
parameters of the lognormal functional form. For each separate method, the values
of the width, $\sigma$, in the two SDP fields are remarkably similar. The average
values, $\langle \sigma \rangle = 1.9\pm0.8$ and $1.8\pm0.6$ $[\ln\,M_\odot]$
for $\ell=30^{\circ}$ and $\ell=59^{\circ}$, respectively, are also very similar.  
Like $M_{\rm inf}$, the value of the characteristic mass, $\mu$, is very different 
in the two regions. \\
\item The similarity of $\alpha$ and $\sigma$ on one side, and the difference of $M_{\rm inf}$
and $\mu$ on the other, show that the CMFs of the two SDP fields have very similar
shapes but different mass scales which, according to our simulations, 
cannot be explained by distance effects alone. This represents an {\it evidence
that the overall process of star formation in the two regions
is very different}. \\
\item The similarity of the shape of the CMF in the two SDP regions suggests 
that if the evolution toward the IMF of high-mass clumps
($\ell=30^{\circ}$) is different from those of low- and
intermediate-mass clumps ($\ell=59^{\circ}$), it leaves
no trace on the shape of the CMF. \\
\item The width of the IMF is narrower than the measured values of $\sigma$ in the
two SDP fields.  This suggests that an additional mass selection occurs in later 
stages of gravitational collapse. \\

\end{itemize}

\appendix

\section{Normalization of the lognormal Mass Function}
\label{sec:normlogn}

By applying the normalization condition (\ref{eq:norm}) to the lognormal MF in
Eq.~(\ref{eq:logn}), we get:
\begin{eqnarray*}
\int_{\ln(M_{\rm inf})}^{\ln(M_{\rm sup})}  \frac{A_{ln}}{\sqrt{2\pi}\sigma} \,
\exp \left [ - \frac{(\ln M - \mu)^2}{2 \sigma^2} \right ] {\rm d}\ln\,M & = & \\ 
& = & N_{\rm tot}
\end{eqnarray*}
which by changing the variable of integration can be easily transformed into:
\begin{equation}
\frac{A_{ln}}{\sqrt{\pi}} \, \int_{x_{\rm inf}}^{x_{\rm sup}} \exp(-x^2) {\rm d} x 
= N_{\rm tot}
\label{eq:Aln1}
\end{equation}
where we have defined the variable $x(M)=(\ln M - \mu)/(\sqrt{2} \sigma)$, 
and $x_{\rm inf} = x(M_{\rm inf})$ and $x_{\rm sup} = x(M_{\rm sup})$.  
Then, by using the following relation 
\begin{eqnarray*}
\int_{0}^{x_1} \exp(-x^2) {\rm d} x +
\int_{x_1}^{x_2} \exp(-x^2) {\rm d} x & + &  \\
+ \int_{x_2}^{\infty} \exp(-x^2) {\rm d} x = 
\int_{0}^{\infty} \exp(-x^2) {\rm d} x & = & \frac{\sqrt{\pi}}{2}
\end{eqnarray*}
which converts into:
\begin{eqnarray}
\int_{x_1}^{x_2} \exp(-x^2) {\rm d} x & = & \frac{\sqrt{\pi}}{2} \left[ 
1 - {\rm erf}(x_1) - {\rm erfc}(x_2) \right] =   \\ \nonumber
& = & \frac{\sqrt{\pi}}{2} \left[ 
{\rm erfc}(x_1) - {\rm erfc}(x_2) \right]
\label{eq:Aln2}
\end{eqnarray}
we can write Eq.~(\ref{eq:Aln1}) as:
\begin{equation}
\frac{A_{ln}}{2} \, \left[ 1 - {\rm erf}(x_{\rm inf}) - {\rm erfc}(x_{\rm sup}) \right] = N_{\rm tot}
\label{eq:Aln3}
\end{equation}
where the ${\rm erf}$ and ${\rm erfc}$ are defined as:
\begin{eqnarray*}
& & {\rm erf}(x) = \frac{2}{\sqrt{\pi}} \int_{0}^{x} \exp(-t^2) {\rm d} t  \,\,\, {\rm and} \\
& & {\rm erfc}(x) = 1 - {\rm erf}(x) = \frac{2}{\sqrt{\pi}} \int_{x}^{\infty} \exp(-t^2) {\rm d} t \, .
\end{eqnarray*}
Finally, we can write the normalization constant, $A_{ln}$, as:
%
\begin{eqnarray}
& & A_{ln} = 2 N_{\rm tot} \times \\
& & \left[ {\rm erfc} \left( \frac{(\ln M_{\rm inf} - \mu)^2}{2 \sigma^2} \right) - 
{\rm erfc} \left( \frac{(\ln M_{\rm sup} - \mu)^2}{2 \sigma^2} \right)  \right]^{-1} . \nonumber
\label{eq:Aln}
\end{eqnarray}
%

\section{Estimating the $M_{\rm inf}$ and $M_{\rm sup}$ parameters }
\label{sec:lognparameters}

The most common and easiest ways of choosing the $M_{\rm inf}$ and $M_{\rm sup}$ 
parameters (for both the power-law and lognormal functional forms) 
are either to take the minimum (above the completeness limit) and maximum values 
of the mass range in the dataset, or to plot a histogram of $M$ and choose ${M}_{\rm inf}$ and
${M}_{\rm sup}$ based on a (arbitrary) threshold occupancy for the bins. 
A more robust approach is desirable. In the method we use here, we choose the values of
${M}_{\rm inf}$ and ${M}_{\rm sup}$ that make the probability distributions of the
observed data and the best-fit lognormal model as similar as possible in the range
$\left [ {M}_{\rm inf}, {M}_{\rm sup} \right]$ (\citealp{clauset2009}).

There are a variety of measures for quantifying the distance between two probability distributions,
and following \citet{clauset2009} we choose the Kolmogorov-Smirnov (or KS) statistics,
which is simply the maximum distance between the CCDFs (see Section~\ref{sec:defs}) 
of the observed data, $P_c^{\rm obs}(M)$, and the fitted model, $P_c^{\rm mod}(M)$:
\begin{equation}
D = {\rm max}_{_{ \left[ M_{\rm inf} \leq M \leq M_{\rm sup} \right ] } }
| P_c^{\rm obs}(M) - P_c^{\rm mod}(M) |
\label{eq:ks}
\end{equation}
The procedure that implements this method thus follows four basic steps:%
\begin{itemize}
  \item{(1)} choose $M_{\rm inf}$ and $M_{\rm sup}$ from selected (arbitrary) intervals;
  \item{(2)} calculate the MLE values of $\mu$ and $\sigma$ using Powell's method;
  \item{(3)} apply the KS statistic to the interval $\left [ {M}_{\rm inf},
             {M}_{\rm sup} \right]$ and estimate $D$;
  \item{(4)} go back to step (1) and keep exploring the $(M_{\rm inf}, M_{\rm sup})$ space.
\end{itemize}
At the end of this procedure, we choose the values of $\mu$, $\sigma$, 
${M}_{\rm inf}$ and ${M}_{\rm sup}$ that minimizes $D$. $N_{\rm tot}$ is thus
the total number of objects with mass in the range $\left [ {M}_{\rm inf},
{M}_{\rm sup} \right]$.


\begin{acknowledgements}
We thank Sean Carey for making the 24\,$\mu$m catalog avialable to us
prior to its official release.
\end{acknowledgements}

\bibliographystyle{aa} 
\bibliography{refs}    

\end{document}